\documentclass[sigplan]{acmart}
\pdfoutput=1 
\def\BibTeX{{\rm B\kern-.05em{\sc i\kern-.025em b}\kern-.08emT\kern-.1667em\lower.7ex\hbox{E}\kern-.125emX}}
    
\copyrightyear{2019} 
\acmYear{2019} 
\setcopyright{acmcopyright}
\acmConference[ASPLOS'19]{2019 Architectural Support for Programming Languages and Operating Systems}{April 13--17, 2019}{Providence, RI, USA}
\acmPrice{15.00}
\acmDOI{http://dx.doi.org/10.1145/XXXXXX.XXXXXX}
\acmISBN{ISBN 978-1-4503-6240-5/19/04} 
\usepackage[normalem]{ulem}
\usepackage{soul}
\usepackage{booktabs}
\usepackage{bbding}
\usepackage{pifont}

\usepackage{graphicx}
\usepackage{multirow}
\usepackage{siunitx}

\newcommand{\ignore}[1]{}
\newcommand{\Fig}[1]{Figure~\ref{#1}}

\newcommand{\shrinkBeforeFigureCaption}{\vspace{-0.0cm}}
\newcommand{\shrinkAfterFigure}{\vspace{-0.0cm}}

\newif\ifsubmit
\submitfalse
\ifsubmit
    \newcommand{\aayush}[1]{}
    \newcommand{\dejan}[1]{}
    \newcommand{\geoffrey}[1]{}
    \newcommand{\izzat}[1]{}
    \newcommand{\johnpaul}[1]{}
    \newcommand{\kaushik}[1]{}
    \newcommand{\paolo}[1]{}
    \newcommand{\sai}[1]{}
    \newcommand{\wenmei}[1]{}
    \newcommand{\todo}[1]{}
    \newcommand{\tocite}[1]{}
    
    \newcommand{\postreviewrm}[1]{}
    \newcommand{\mayberm}[1]{}
\else
    \newcommand{\aayush}[1]{[{\color{cyan}AA: #1}]}
    \newcommand{\dejan}[1]{[{\color{cyan}DM: #1}]}
    \newcommand{\geoffrey}[1]{[{\color{cyan}GN: #1}]}
    \newcommand{\izzat}[1]{[{\color{cyan}IE: #1}]}
    \newcommand{\johnpaul}[1]{[{\color{cyan}JPS: #1}]}
    \newcommand{\kaushik}[1]{[{\color{cyan}KR: #1}]}
    \newcommand{\paolo}[1]{[{\color{cyan}PF: #1}]}
    \newcommand{\sai}[1]{[{\color{cyan}SC: #1}]}
    \newcommand{\wenmei}[1]{[{\color{cyan}WMH: #1}]}
    \newcommand{\todo}[1]{[{\color{red}TODO: #1}]}
    \newcommand{\tocite}[1]{[{\color{red}CITE: #1}]}
    
    \newcommand{\postreviewrm}[1]{{\color{red}\sout{#1}}}
    \newcommand{\mayberm}[1]{{\color{gray}\sout{#1}}}
\fi

\begin{document}

%
% The "title" command has an optional parameter, allowing the author to define a "short title" to be used in page headers.
\title{PUMA: A Programmable Ultra-efficient Memristor-based Accelerator\\for Machine Learning Inference}

%
% The "author" command and its associated commands are used to define the authors and their affiliations.
% Of note is the shared affiliation of the first two authors, and the "authornote" and "authornotemark" commands
% used to denote shared contribution to the research.

\author{Aayush Ankit}
\affiliation{
  \institution{Purdue University,\\Hewlett Packard Enterprise}
}

\author{Izzat El Hajj}
\authornote{Work done while at University of Illinois at Urbana-Champaign}
\affiliation{
  \institution{American University of Beirut}
}

\author{Sai Rahul Chalamalasetti}
\affiliation{
  \institution{Hewlett Packard Enterprise}
}

\author{Geoffrey Ndu}
\affiliation{
  \institution{Hewlett Packard Enterprise}
}

\author{Martin Foltin}
\affiliation{
  \institution{Hewlett Packard Enterprise}
}

\author{R. Stanley Williams}
\affiliation{
  \institution{Hewlett Packard Enterprise}
}

\author{Paolo Faraboschi}
\affiliation{
  \institution{Hewlett Packard Enterprise}
}

\author{Wen-mei Hwu}
\affiliation{
  \institution{University of Illinois at Urbana-Champaign}
}

\author{John Paul Strachan}
\affiliation{
  \institution{Hewlett Packard Enterprise}
}
\author{Kaushik Roy}
\affiliation{
  \institution{Purdue University}
}

\author{Dejan S Milojicic}
\affiliation{
  \institution{Hewlett Packard Enterprise}
}

%
% By default, the full list of authors will be used in the page headers. Often, this list is too long, and will overlap
% other information printed in the page headers. This command allows the author to define a more concise list
% of authors' names for this purpose.
\renewcommand{\shortauthors}{Ankit et al.}

%
% The abstract is a short summary of the work to be presented in the article.

\begin{abstract}

Memristor crossbars are circuits capable of performing analog matrix-vector multiplications, overcoming the fundamental energy efficiency limitations of digital logic.
They have been shown to be effective in special-purpose accelerators for a limited set of neural network applications.

We present the Programmable Ultra-efficient Memristor-based Accelerator (PUMA) which enhances memristor crossbars with general purpose execution units to enable the acceleration of a wide variety of Machine Learning (ML) inference workloads.
PUMA's microarchitecture techniques exposed through a specialized Instruction Set Architecture (ISA) retain the efficiency of in-memory computing and analog circuitry, without compromising programmability.

We also present the PUMA compiler which translates high-level code to PUMA ISA.
The compiler partitions the computational graph and optimizes instruction scheduling and register allocation to generate code for large and complex workloads to run on thousands of spatial cores.

We have developed a detailed architecture simulator that incorporates the functionality, timing, and power models of PUMA's components to evaluate performance and energy consumption.
A PUMA accelerator running at 1 GHz can reach area and power efficiency of 577~GOPS/s/mm\textsuperscript{2} and 837~GOPS/s/W, respectively.
Our evaluation of diverse ML applications from image recognition, machine translation, and language modelling (5M-800M synapses) shows that PUMA achieves up to 2,446$\times$ energy and 66$\times$ latency improvement for inference compared to state-of-the-art GPUs.
Compared to an application-specific memristor-based accelerator, PUMA incurs small energy overheads at similar inference latency and added programmability.

\end{abstract}


\copyrightyear{2019} 
\acmYear{2019} 
\setcopyright{acmcopyright}
\acmConference[ASPLOS '19]{2019 Architectural Support for Programming Languages and Operating Systems}{April 13--17, 2019}{Providence, RI, USA}
\acmBooktitle{2019 Architectural Support for Programming Languages and Operating Systems (ASPLOS '19), April 13--17, 2019, Providence, RI, USA}
\acmPrice{15.00}
\acmDOI{10.1145/3297858.3304049}
\acmISBN{978-1-4503-6240-5/19/04}

%
% Keywords. The author(s) should pick words that accurately describe the work being
% presented. Separate the keywords with commas.

%
% A "teaser" image appears between the author and affiliation information and the body 
% of the document, and typically spans the page. 
%  \Description{Enjoying the baseball game from the third-base seats. Ichiro Suzuki preparing to bat.}

%
% This command processes the author and affiliation and title information and builds
% the first part of the formatted document.
\maketitle

\section{Introduction}\label{sec:intro_new}

% MOTIVATE MACHINE LEARNING ACCELERATORS
General-purpose computing systems have benefited from scaling for several decades, but are now hitting an energy wall.
This trend has led to a growing interest in domain-specific architectures.
Machine Learning (ML) workloads in particular have received tremendous attention because of their pervasiveness in many application domains and high performance demands.
Several architectures have been proposed, both digital~\cite{chen2016eyeriss,chen2014dadiannao,farabetneuflow2010,Jouppi:2017:IPA:3079856.3080246,liu2016cambricon,reagen2016minerva} and mixed digital-analog using memristor crossbars~\cite{cheng2017time,chi2016prime,liu2015reno,shafiee2016isaac,song2017pipelayer}.

% MACHINE LEARNING WITH MEMRISTOR
ML workloads tend to be data-intensive and perform a large number of Matrix Vector Multiplication (MVM) operations.
Their execution on digital CMOS hardware is typically characterized by high data movement costs relative to compute~\cite{han2015learning}.
To overcome this limitation, memristor crossbars can store a matrix with high storage density and perform MVM operations with very low energy and latency~\cite{alibart2013pattern,Burr2015IEDM,Hu2018MNIST,prezioso2015,sheridan2017sparse,Yu2016_IEDM_16Mb}.
Each crosspoint in the crossbar stores a multi-bit value in one memristor device, which enables high storage density~\cite{waser2009redox}.
Upon applying an input voltage at the crossbar's rows, we get the MVM result as output current at the crossbar's columns based on Kirchhoff's law.
A crossbar thus performs MVM in one computational step -- including \textit{O(n\textsuperscript{2})} multiplications and additions for an $n \times n$ matrix -- which typically takes many steps in digital logic.
It also combines compute and storage in a single device to alleviate data movement, thereby providing intrinsic suitability for data-intensive workloads~\cite{chi2016prime,shafiee2016isaac}.

% MEMRISTOR ACCELERATOR RESEARCH AND LIMITATIONS
Memristor crossbars have been used to build special-purpose accelerators for Convolutional Neural Networks (CNN) and Multi Layer Perceptrons (MLP)~\cite{chi2016prime,liu2015reno,shafiee2016isaac}, but these designs lack several important features for supporting general ML workloads.
% (1) State machines are not scalable
First, each design supports one or two types of neural networks, where layers are encoded as state machines.
This approach is not scalable to a larger variety of workloads due to increased decoding overhead and complexity.
%Typical applications are based on multiple ML algorithms.
% (2) Missing linear and transcendental operations
Second, existing accelerators lack the types of operations needed by general ML workloads.
\ignore{
For example, Long Short-Term Memory (LSTM) is a sequence processing algorithm widely used in language processing and speech-text engines~\cite{hochreiter1997long}.
It requires multiple vector linear and transcendental functions, in addition to MVM, which cannot be executed on crossbars and are not supported by existing designs.
}
For example, Long Short-Term Memory (LSTM)~\cite{hochreiter1997long} workloads require multiple vector linear and transcendental functions which cannot be executed on crossbars efficiently and are not supported by existing designs.
% (3) Data movement
Third, existing designs do not provide flexible data movement and control operations to capture the variety of access and reuse patterns in different workloads.
Since crossbars have high write latency~\cite{hu2016dac}, they typically store constant data while variable inputs are routed between them in a spatial architecture.
This data movement can amount to a significant portion of the total energy consumption which calls for flexible operations to optimize the data movement.
\ignore{
% (4) ISAAC deep pipeline, PRIME Precision
Fourth, some designs use deep pipelines~\cite{shafiee2016isaac} which are not suitable for general workloads with control flow.
Finally, other designs~\cite{chi2016prime} make optimistic assumptions about precision which may not be suitable for all types of workloads.
}

To address these limitations, we present PUMA, a Programmable Ultra-efficient Memristor-based Accelerator.
PUMA is a spatial architecture designed to preserve the storage density of memristor crossbars to enable mapping ML applications using on-chip memory only.
It supplements crossbars with an instruction execution pipeline and a specialized ISA that enables compact representation of ML workloads with low decoder complexity.
It employs temporal SIMD units and a ROM-Embedded RAM~\cite{lee2013area} for area-efficient linear and transcendental vector computations.
It includes a microarchitecture, ISA, and compiler co-designed to optimize data movement and maximize area and energy efficiency.
To the best of our knowledge, PUMA is the first programmable and general-purpose ML inference accelerator built with hybrid CMOS-memristor technology.

% CHALLENGE IN GENERALITY
A na\"ive approach to generality is not viable because of the huge disparity in compute and storage density between the two technologies.
CMOS digital logic has an order of magnitude higher area requirement than a crossbar for equal output width ($\sim$20$\times$, see Table~\ref{tab:hardwaretab}).
Moreover, a crossbar's storage density (2-bit cells) is $160MB/mm\textsuperscript{2}$, which is at least an order of magnitude higher than SRAM (6T, 1-bit cell)~\cite{shafiee2016isaac}.
A 90mm\textsuperscript{2} PUMA node can store ML models with up to 69MB of weight data.
Note that the PUMA microarchitecture, ISA, and compiler are equally suitable to crossbars made from emerging technologies other than memristors such as STT-MRAM~\cite{sengupta2016proposal}, NOR Flash~\cite{guo2017fast}, etc.

% PUMA SUMMARY
We make the following contributions:
\begin{itemize} \setlength\itemsep{-0.02in}
    \item A programmable and highly efficient architecture exposed by a specialized ISA for scalable acceleration of a wide variety of ML applications using memristor crossbars.
    \item A complete compiler which translates high-level code to PUMA ISA, enabling the execution of complex workloads on thousands of spatial cores.
    \item A detailed simulator which incorporates functionality, timing, and power models of the architecture.
    \item An evaluation across ML workloads showing that PUMA can achieve promising performance and energy efficiency compared to state-of-the-art CPUs, GPUs, TPU, and application-specific memristor-based accelerators.
\end{itemize}
Our simulator and compiler have been open-sourced to enable further research on this class of accelerators.

\section{Workload Characterization}\label{sec:workload}

This section characterizes different ML inference workloads with a batch size of one.
The characteristics are summarized in Table~\ref{tab:workload-analysis}.
\ignore{
Common characteristics across workloads include
    dominance of MVM operations,
    high data parallelism, and
    use of nonlinear operations.
Characteristics that differ across workloads include
    use of linear and transcendental operations,
    weight and input data reuse,
    resource boundedness, and
    memory access pattern.
}
The section's objective is to provide insights on the suitability of memristor crossbars for accelerating ML workloads and highlight implications on the proposed architecture.

\subsection{Multi-Layer Perceptron (MLP)}\label{sec:mlp}
MLPs are neural networks used in common classification tasks such as digit-recognition, web-search, etc.~\cite{cirecsan2010deep, Jouppi:2017:IPA:3079856.3080246}.
Each layer is fully-connected and applies a nonlinear function to the weighted-sum of outputs from the previous layer.
The weighted-sum is essentially an MVM operation.
Equation \ref{eqn:mlp} shows the computations in a typical MLP ($act$ is nonlinear):
% MLP equation
\vspace{-2mm}
\begin{equation}\label{eqn:mlp}
    \small
    O[y] = act(B[y] + \sum_{x=0}^{n-1}I[x] \times W[x][y])
    \vspace{-2mm}
\end{equation}

\ignore{
MLPs are the simplest type of ML workload discussed in this section.
They capture all the common features across these workloads, namely
}
MLPs are simple, capturing the features common across the ML workloads we discuss:
    dominance of MVM operations,
    high data parallelism, and
    use of nonlinear operations.

\subsubsection{Dominance of MVM}
MVM operations are $O(n^2)$ in space and computational complexity, whereas the nonlinear operations are $O(n)$, where $n$ is the matrix dimension (layer size).
MVMs are therefore the dominant operation in MLPs (and other ML workloads).
This property makes memristor crossbars suitable for acceleration since they perform analog MVMs with low energy/latency.

\begin{table}[t]
\centering
    \caption{Workload Characterization}\label{tab:workload-analysis}
    \vspace{-0.4cm}
    \centering
    \resizebox{0.75\columnwidth}{!}{
        
\begin{tabular}{|l|c|c|c|}
    \hline
    \multicolumn{1}{|c|}{\textbf{Characteristic}} & \textbf{MLP} & \textbf{LSTM} & \textbf{CNN} \\
    \hline
    \textbf{Dominance of MVM} & Yes & Yes & Yes \\
    \hline
    \textbf{High data parallelism} & Yes & Yes & Yes \\
    \hline
    \textbf{Nonlinear operations} & Yes & Yes & Yes \\
    \hline
    \textbf{Linear operations} & No & Yes & No \\
    \hline
    \textbf{Trancendental operations} & No & Yes & Yes \\
    \hline
    \textbf{Weight data reuse} & No & Yes & Yes \\
    \hline
    \textbf{Input data reuse} & No & No & Yes \\
    \hline
    \textbf{Bounded resource} & Memory & Memory & Compute \\
    \hline
    \textbf{Sequential access pattern} & Yes & Yes & No \\
    \hline
\end{tabular}

    }
    \vspace{-0.5cm}
\end{table}

\subsubsection{High data parallelism}
MLPs (and other ML workloads) have massive amounts of data parallelism.
Moreover, practical model sizes are larger than the typical on-chip storage that can be provided by SRAM.
For this reason, CMOS implementations suffer from costly DRAM accesses which are particularly taxing due to the absence of data reuse to amortize them.
On the other hand, crossbars have extremely high area efficiency which allows deploying many of them on a single chip.
Doing so not only captures the high data parallelism in these workloads, but it also provides high storage density to fit models on-chip and bypass costly DRAM accesses.

\subsubsection{Nonlinear operations}\label{sec:mlp-nonlinear}
In addition to MVM operations, MLPs (and other ML workloads) perform nonlinear vector operations (e.g., ReLU).
Since these operations cannot be performed in crossbars, an implication on the architecture is the need to provide digital functional units to support them.
Such functional units consume significantly more area ($\sim$20$\times$) than crossbars for equal output width (see Table~\ref{tab:hardwaretab}).
The challenge is to size these units appropriately to provide sufficient throughput without offsetting crossbar area/energy efficiency.

\subsection{Long Short-Term Memory (LSTM)}
LSTMs are the state-of-the-art technique for sequence processing tasks like speech processing, language modelling, etc.~\cite{hochreiter1997long}.
Each layer is fully connected and performs linear and nonlinear operations on the weighted-sum of outputs and the previous state.
These operations translate into two MVMs followed by multiple (typically four) vector arithmetic operations and (typically four) nonlinear functions.
Equations \ref{eqn:lstm1} to \ref{eqn:lstm3} show the computations in a typical LSTM:
% LSTM equation
\vspace{-3mm}
\begin{equation}\label{eqn:lstm1}
\small
    F\textsubscript{t}[y] = act(B[f] + \sum_{x=0}^{n-1}(H,I)[x] \times W\textsubscript{f}[x][y])
\end{equation}
\vspace{-6mm}
\begin{equation}\label{eqn:lstm2}
\small
    C\textsubscript{t}[y] = \sum_{x=0}^{n-1} (f\textsubscript{t}[y] \times C\textsubscript{t-1}[y] + g\textsubscript{t}[y] \times Cp\textsubscript{t-1}[y])
\end{equation}
\vspace{-4mm}
\begin{equation}\label{eqn:lstm3}
\small
    H\textsubscript{t}[y] = \sum_{x=0}^{n-1} (h\textsubscript{t}[y] \times C\textsubscript{t}[y])
\end{equation}
To the best of our knowledge, PUMA is the first memristor-based accelerator demonstrated with LSTMs.

\subsubsection{Linear and transcendental operations}\label{sec:lstm-linear-trancend}
Unlike MLPs, LSTMs also perform linear vector operations.
Moreover, the typical nonlinear vector operations in LSTMs are transcendental (e.g. tanh, sigmoid).
Supporting these operations has the same implication on the architecture as discussed in Section~\ref{sec:mlp-nonlinear} for nonlinear operations.
Transcendental operations are particularly challenging due to their high complexity.

\subsubsection{Weight reuse}
Another key distinction of LSTMs compared to MLPs is data reuse.
LSTM inputs consist of a sequence of vectors processed across multiple time-steps with the same weights.
This feature benefits CMOS architectures by amortizing DRAM accesses for loading weights, but is not advantageous to memristor crossbars.
That said, the scope of weight reuse in LSTMs is only over a few inputs so the workload remains memory-bound.
It still suffers in CMOS hardware from insufficient amortization of DRAM accesses.

\subsection{Convolutional Neural Network (CNN)}\label{sec:workload-cnn}
CNNs are widely used for image recognition and classification~\cite{krizhevsky2012imagenet}.
They typically include several convolutional, pooling, and response normalization layers. 
A convolution layer consists of weight kernels strided across the input image in a sliding window fashion.
It exhibits a non-sequential memory access pattern since a window of the input consists of parts of the input image from different rows.
Equation~\ref{eqn:cnn} shows the computations in a typical convolutional layer of a CNN:
% CNN equation
\vspace{-2mm}
\begin{equation}\label{eqn:cnn}
\small
\begin{split}
    O[m][x][y] = act(B[m] + \\
    \sum_{i=0}^{R-1}\sum_{j=0}^{S-1}\sum_{k=0}^{C-1} I[k][Ux+i][Uy+j]\times & W[m][k][i][j])
\end{split}
\end{equation}
\vspace{-4mm}

\subsubsection{Input reuse and compute intensity}\label{sec:workload-input-compute}
Convolution layers exhibit both weight and input data reuse.
They can be mapped to matrix-matrix multiplications which successively apply weight kernels on different input windows.
Matrix-matrix multiplications are compute-bound which makes them well-suited for CMOS hardware since there is enough data reuse to amortize DRAM access cost.
However, memristor crossbars can still perform well on matrix-matrix operations by treating them as successive MVMs.
An implication on architecture is the opportunity to take advantage of input reuse to minimize data movement within the chip.
Another implication is that iterating over inputs creates the need for control flow to represent the workload compactly without code bloat.

\subsubsection{Non-sequential access}\label{sec:workload-access-pattern}
Unlike MLPs and LSTMs, CNNs exhibit non-sequential accesses due to the way inputs are traversed as well as the behavior of non-convolutional layers such as pooling and response-normalization.
An implication on the architecture is the need to support fine-grain/random access to memory, which is not needed for MLPs and LSTMs where it is sufficient to access data at the granularity of the input/output vectors to each layer.

\subsection{Other ML Workloads}

Other workloads, both supervised and unsupervised, can be represented using a combination of the patterns in the three applications in this section.
\textit{Logistic Regression}~\cite{agresti2002logistic} and \textit{Linear Regression}~\cite{neter1996applied} compute weighted-sums which are passed to activation functions to generate probabilities and continuous values respectively. 
\textit{Support Vector Machine (SVM)}~\cite{furey2000support} and \textit{Recommender Systems}~\cite{resnick1997recommender} compute weighted-sums followed by nonlinear functions.
Their computations are similar to MLP.
\textit{Recurrent Neural Networks (RNNs)}~\cite{mikolov2010recurrent} used for sequence processing compute weighted-sums on input and previous state.
They are similar to LSTMs but without vector operations. 
\textit{Generative Adversarial Networks (GANs)} are composed of two neural networks (MLP, LSTM, CNN, etc.) which compete to reach equilibrium~\cite{goodfellow2014generative}.
\textit{Restricted Boltzmann Machines (RBM)}~\cite{sutskever2009recurrent} and \textit{Boltzmann Machines (BM)}~\cite{tanaka1998mean} are commonly used in unsupervised learning tasks for energy-minimization. 
While RBM involves weighted-sums of previous state and inputs, BM uses inputs only.
Their computations have similarities to MLPs and LSTMs as well.

\section{Core Architecture}\label{sec:architecture}

We propose a programmable architecture and ISA design that leverage memristor crossbars for accelerating ML workloads.
PUMA is a spatial architecture organized in three-tiers: cores, tiles, and nodes.
Cores consist of analog crossbars, functional units, and an instruction execution pipeline.
Tiles consist of multiple cores connected via a shared memory.
Nodes consist of multiple tiles connected via an on-chip network.
Subsequently, nodes can be connected together via a chip-to-chip interconnect for large-scale execution.

\begin{figure}[t]
  \centering
  \includegraphics[width=0.95\columnwidth]{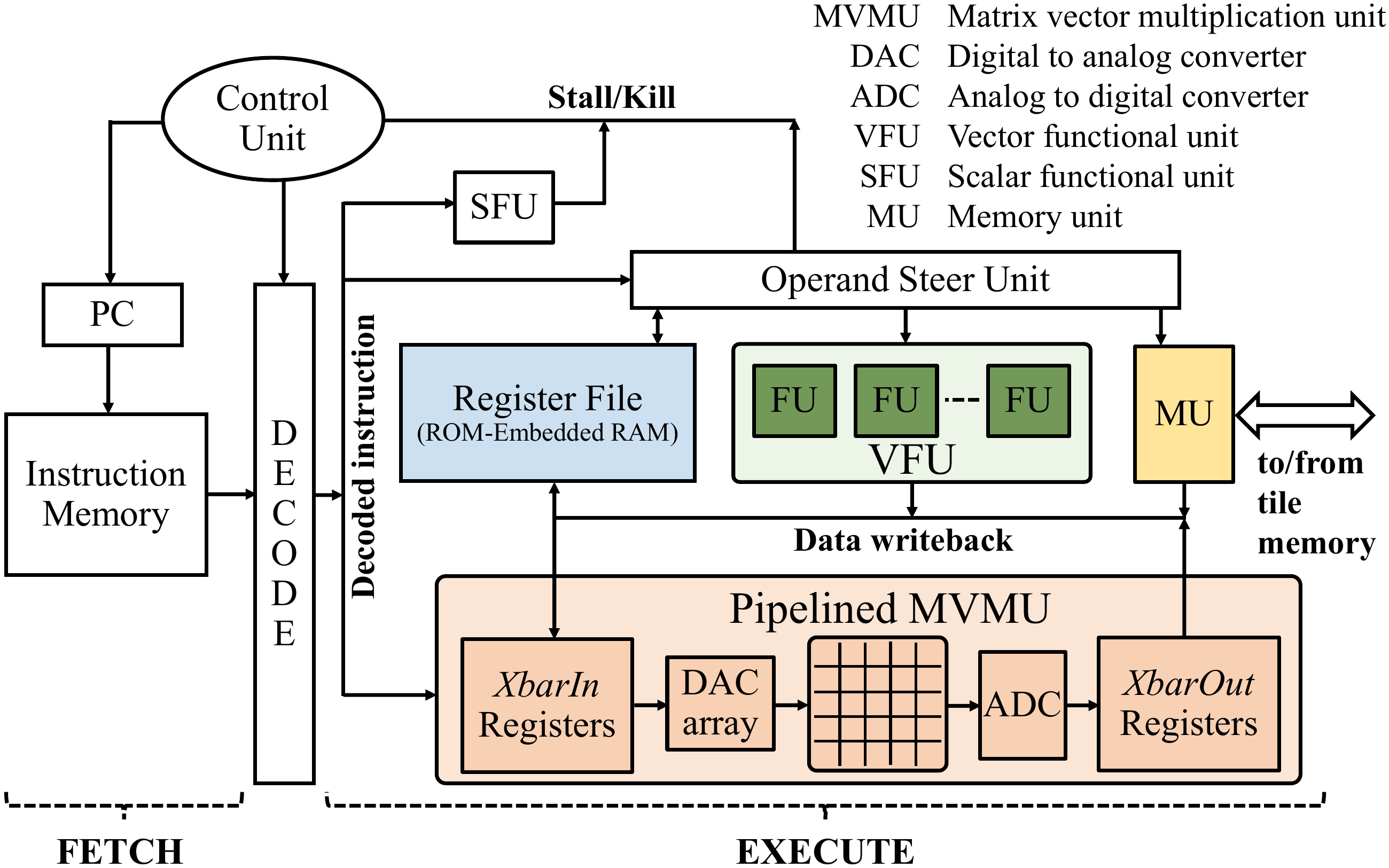}
  \vspace{-0.3cm}
  \caption{Core Architecture}\label{fig:core-architecture}
  \vspace{-0.3cm}
\end{figure}

While this hierarchical organization is common in related work~\cite{chen2014dadiannao,shafiee2016isaac}, our key contributions lie in the core architecture (this section) and tile architecture (Section~\ref{sec:tile}) that bring programmability and generality to memristor crossbars without compromising their energy and area efficiency.
An overview of the core architecture is shown in \Fig{fig:core-architecture}.
The following subsections discuss the components of the core architecture and the insights behind their design.

\begin{table*}[t]
\centering
    \caption{Instruction Set Architecture Overview}\label{tab:isa_over}
    \vspace{-0.4cm}
    \centering
    \resizebox{0.75\textwidth}{!}{
        \begin{tabular}{|l|l|l|l|}
    \hline
    \multicolumn{1}{|c|}{\textbf{Category}} & \multicolumn{1}{|c|}{\textbf{Instruction}} & \multicolumn{1}{|c|}{\textbf{Description}} & \multicolumn{1}{|c|}{\textbf{Operands}}\\
    \hline
    \multirow{6}{*}{Compute} & MVM & Matrix-Vector Multiplication & mvm, mask, -, filter, stride, -, -\\
    \cline{2-4}
            & \multirow{3}{*}{ALU} & Vector arithmetic/logical (add, subtract, multiply, divide, shift, and, or, invert) & \multirow{3}{*}{alu, aluop, dest, src1, src2, src3, vec-width} \\
            &     & Vector non-linear (relu, sigmoid, tanh, log, exponential)  &                                                \\
            &     & Other             (random vector, subsampling, min/max)    &                                                \\
    \cline{2-4}
            & ALUimm & Vector arithmetic immediate (add, subtract, multiply, divide)        & alui, aluop, dest, src1, immediate, vec-width  \\
    \cline{2-4}
            & ALUint & Scalar arithmetic (add, subtract) - Compare (equal, greater than, not equal)  & alu-int, aluop, dest, src1, src2, -, -         \\
    \hline
    Intra-Core & set & Register initialization                   & set, -, dest, immediate, -, -                  \\
    \cline{2-4}
    Data Movement & copy & Data movement between different registers & copy, -, dest, src1, -, ld-width, vec-width   \\
    \hline
    Intra-Tile & load & Load data from shared memory                   & load, -, dest, immediate, -, -                  \\
    \cline{2-4}
    Data Movement & store & Store data to shared memory & store, -, dest, src1, count, st-width, vec-width   \\
    \hline
    Intra-Node & send    &  Send data to tile &  send, memaddr, fifo-id, target, send-width, vec-width \\
    \cline{2-4}
    Data Movement & receive &  Receive data from tile & receive, memaddr, fifo-id, count, rec-width, vec-width \\
    \hline
    \multirow{2}{*}{Control} & jmp & Unconditional jump & jmp, -, -, -, -, pc \\
    \cline{2-4}
        & brn & Conditional jump   & brn, brnop, -, src1, src2, pc \\
    \hline
    
\end{tabular}
    }
    \vspace{-0.3cm}
\end{table*}

\subsection{Instruction Execution Pipeline}\label{sec:core-pipeline}

Existing memristor-based accelerators~\cite{chi2016prime,liu2015reno,shafiee2016isaac} are limited to one or two ML workloads.
They use state machines that can be configured to compose a small set of functional blocks (e.g., convolution block, pooling block, etc.).
While this approach works well when the scope of workloads is small, supporting a larger variety of workloads creates high decoding complexity.
For this reason, our core architecture breaks functionality down to finer-grain instructions and supplements memristor crossbars with an instruction execution pipeline.
Our approach is based on the observation in Section~\ref{sec:workload} that despite the large variety of ML workloads, these workloads share many low-level operations.

The instruction execution pipeline is an in-order pipeline with three stages: fetch, decode, and execute.
Keeping the pipeline simple saves area to avoid offsetting the crossbars' area efficiency.
The ISA executed by the pipeline is summarized in Table~\ref{tab:isa_over}.
Instructions are seven bytes wide.
The motivations for wide instructions are discussed in Sections~\ref{sec:core-vfu} and~\ref{sec:isa-implications}.
The ISA instruction usage is shown in Section~\ref{sec:core-usage}.
More ISA details are discussed in another paper~\cite{ambrosi2018hardware}.

The instruction execution pipeline supports control flow instructions (\textit{jmp} and \textit{brn} in Table~\ref{tab:isa_over}), as motivated in Section~\ref{sec:workload-input-compute}.
It also includes a \textit{Scalar Functional Unit (SFU)} that performs scalar integer operations (\textit{ALUint} in Table~\ref{tab:isa_over}) to support the control flow instructions.

\begin{figure}[t]
  \centering
  \includegraphics[width=0.75\columnwidth]{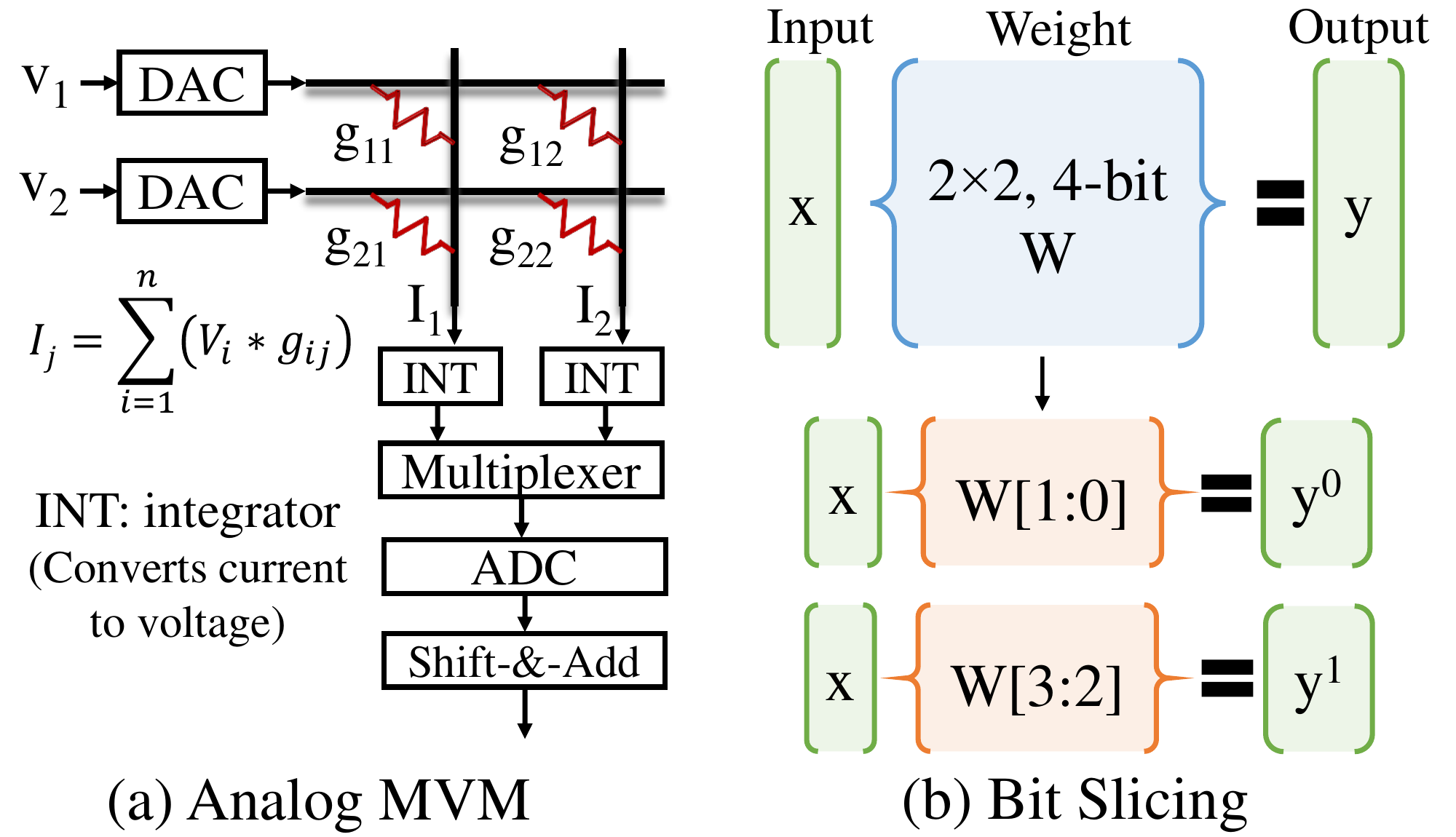}
  \vspace{-0.3cm}
  \caption{MVM with Crossbars}\label{fig:mvm-on-crossbar}
  \vspace{-0.3cm}
\end{figure}

\subsection{Matrix-Vector Multiplication Unit (MVMU)}\label{sec:core-mvmu}

The MVMU (illustrated in \Fig{fig:core-architecture}) consists of memristor crossbars that perform analog MVM operations, and peripherals (DAC/ADC arrays) that interface with digital logic via the \textit{XbarIn} and \textit{XbarOut} registers.
\textit{XbarIn} registers provide digital inputs to the DACs which feed analog voltages to the crossbar.
ADCs convert crossbar output currents to digital values which are stored in the \textit{XbarOut} registers.
This crossbar design is similar to ISAAC~\cite{shafiee2016isaac}.

Figure~\ref{fig:mvm-on-crossbar}(a) shows how memristor crossbars can be used to perform analog MVM operations.
DACs convert the input vector to analog voltages applied at crossbar rows.
The matrix is encoded as conductance states ($g\textsubscript{ij}$) of the devices that constitute the crossbar.
The currents at crossbar columns constitute the MVM result.
They are integrated (converted to voltage) then converted to digital values with ADCs.

\subsubsection{Precision Considerations}

Practically realizable crossbars provide 2-6 bits of precision per device~\cite{Hu2018MNIST}.
We conservatively use 2 bits per device, and realize 16-bit MVM operations by combining 8 crossbars via bit-slicing~\cite{shafiee2016isaac}, illustrated in Figure~\ref{fig:mvm-on-crossbar}(b).
ADCs are reused across columns to save area.
The impact of precision on inference accuracy is evaluated in Section~\ref{sec:results-sweep}.

\subsubsection{Crossbar Co-location and Input Sharing}

Crossbar peripherals have an order of magnitude higher area than the crossbar.
Since all eight 2-bit crossbars of a 16-bit MVM operation are used simultaneously on the same input, we co-locate these 2-bit crossbars on the same core in the same MVMU, which allows us to use the same \textit{XbarIn} registers and DAC array to feed them with minimal routing congestion.
This co-location and input reuse is provided transparently in the ISA, which exposes a full 16-bit MVM operation in a single instruction (\textit{MVM} in Table~\ref{tab:isa_over}).

\subsubsection{Input Shuffling}
As motivated in Section~\ref{sec:workload-input-compute}, ML workloads with sliding window computations typically reuse large portions of the input across successive MVM operations ($\sim$80\% for convolutional layers with filter size 5 and unit stride).
However, reused input values may come at different positions in the input vector.
To avoid moving data around in \textit{XbarIn}, the MVM instruction provides operands (\textit{filter}/\textit{stride} in Table~\ref{tab:isa_over}) that re-route XbarIn registers to different DACs, enabling logical input shuffling without physical data movement.

\subsubsection{Multiple MVMUs per Core}\label{sec:core-mvmu-coalesce}

A core may have multiple MVMUs, in which case it is desirable to activate them in parallel since MVMs are heavy operations.
The in-order pipeline does not capture the Instruction-Level Parallelism (ILP) between MVM instructions automatically.
Instead, the ISA exposes an operand (\textit{mask} in Table~\ref{tab:isa_over}) to allow a single MVM instruction to activate multiple MVMUs at once.
Compiler optimizations that use this operand are discussed in Section~\ref{sec:compiler-coalescing}.

\subsubsection{Crossbar Writes}

PUMA is an inference accelerator, so crossbars are initialized with weights using serial writes at configuration time prior to execution and are not written to throughout execution.
In this sense, PUMA is a spatial architecture where data is routed between crossbars, each crossbar storing a different portion of the model.
Larger models therefore require more area, and may scale to multiple nodes.

\subsection{Vector Functional Unit (VFU)}\label{sec:core-vfu}

The VFU executes linear and nonlinear vector operations (\textit{ALU} and \textit{ALUimm} in Table~\ref{tab:isa_over}), as motivated by Sections~\ref{sec:mlp-nonlinear} and~\ref{sec:lstm-linear-trancend}.
An important aspect of designing vector instructions and the VFU is choosing the vector width.
Since ML workloads have high data parallelism, they execute wide vector operations, which motivates having wide vector instructions.
Wide vector instructions have the benefit of reducing instruction count, and consequently, fetch, decode, and instruction storage overhead.
On the other hand, hardware considerations motivate having narrow VFU vector width to avoid offsetting the area efficiency of crossbars as discussed in Section~\ref{sec:mlp-nonlinear}.

To balance the tension between workloads favoring wide vector width and hardware favoring narrow vector width, we propose a VFU design based on \textit{temporal SIMD}.
Temporal SIMD uses a narrow width VFU to execute wide vectors over multiple cycles.
The vector instruction operands specify the starting address of the vectors in the register file as well as the vector width (\textit{vec-width} in Table~\ref{tab:isa_over}).
The \textit{operand steer unit} holds the decoded instruction and reads the operands from the register file over subsequent cycles to feed the VFU.
The additional \textit{vec-width} operand required by temoral SIMD motivates our wide instruction design.

Provisioning the adequate width for VFUs maintains crossbar area efficiency benefits without the VFU becoming a bottleneck and compromising throughput.
A narrow VFU is possible because typical ML workloads compute $O(n)$ more operations per MVM instruction than per vector instruction.
Section~\ref{sec:results-sweep} evaluates the impact of VFU width on efficiency.

\subsection{Register File}\label{sec:core-rf}
We propose a register file design that uses \textit{ROM-Embedded RAM}~\cite{lee2013area} to accomplish two objectives:
    (1) harboring general purpose registers, and
    (2) providing area-efficient transcendental function evaluations as motivated in Section~\ref{sec:lstm-linear-trancend}.

\subsubsection{Implementing transcendental functions}

\begin{figure}[t]
  \centering
  \includegraphics[width=0.90\columnwidth]{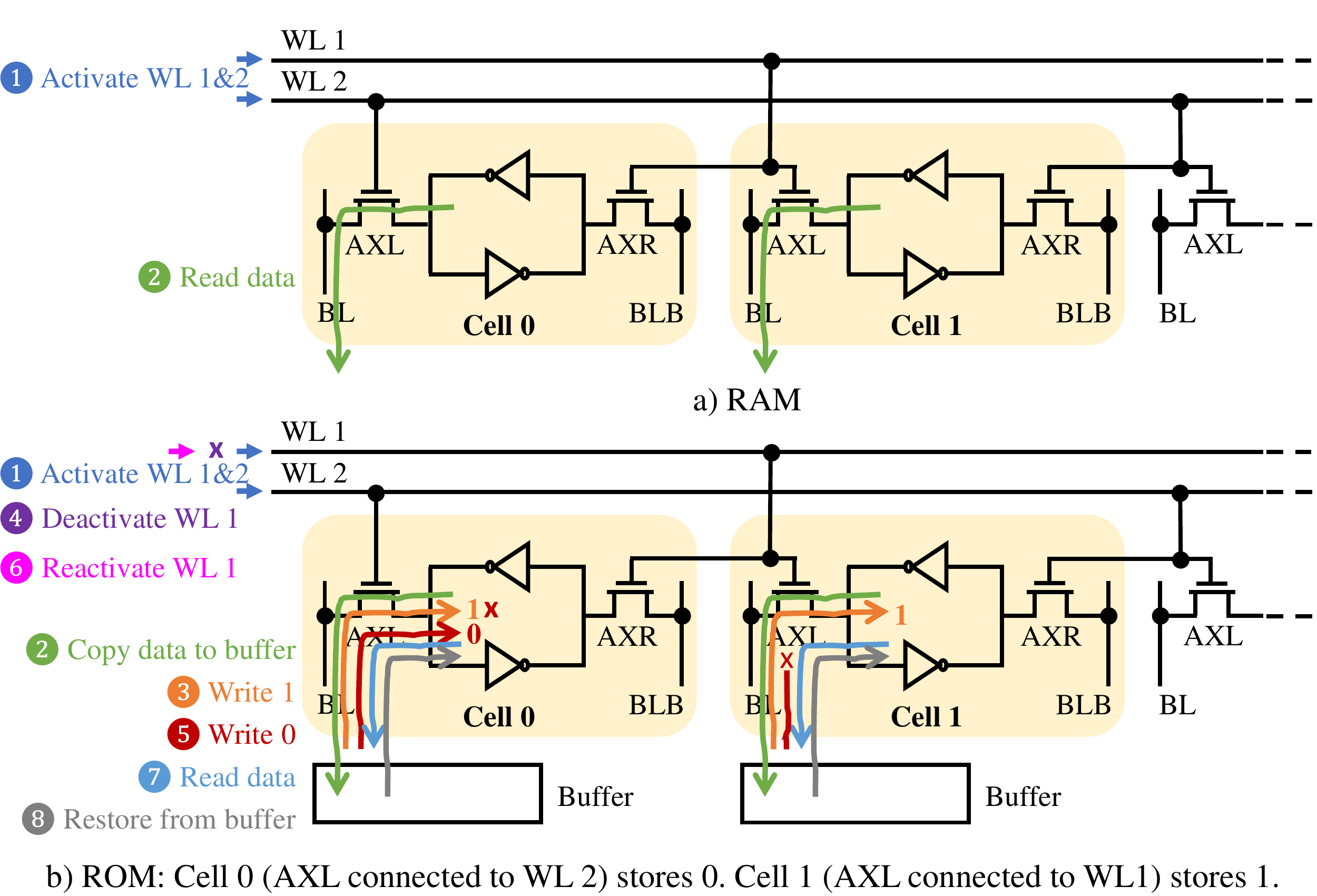}
  \vspace{-0.3cm}
  \caption{ROM-Embedded RAM}\label{fig:emb_rom}
  \vspace{-0.3cm}
\end{figure}

Area-efficient function evaluations are crucial for preserving crossbar storage density.
For this reason, we use a ROM-Embedded RAM structure~\cite{lee2013area} which adds a wordline per row to embed a ROM that is used to store look-up tables without increasing the array area or RAM latency. 
Alternative digital implementations to support transcendental functions are prohibitive due to their high area requirements, especially in light of the large number of transcendental function types used in ML.
Transcendental function evaluations also use temporal SIMD (Section~\ref{sec:core-vfu}) to minimize fetch/decode energy consumption.

\Fig{fig:emb_rom} details the operation of a ROM-Embedded RAM.
In RAM mode, both wordlines ($WL1$ and $WL2$) are activated, followed by precharging or driving the bitlines for read or write operations, respectively (similar to typical SRAM).
In ROM mode, since a ROM access overwrites the RAM data, the first step is to buffer the RAM data.
Subsequently, 1 is written to all cells with both wordlines activated.
Next, 0 is written to all cells while keeping the WL1 deactivated, which writes 0 to a cell only if its AXL is connected to WL2.
Therefore, cells with AXL connected to WL1 and WL2, will store a ROM value of 1 and 0, respectively.
A read with both wordlines activated is done to retrieve the ROM data, followed by restoring the original RAM contents.

\subsubsection{Sizing the register file}

The register file enables buffering data in general purpose registers to avoid higher-cost access to shared memory.
However, if the register file were too large, it would degrade core storage density.
A key point to observe is that the majority of ML kernels are such that data is consumed within 1-2 instructions after being produced.
This property is preserved via proper instruction scheduling by the compiler to reduce register pressure (Section~\ref{sec:compiler-instruction-pressure}).
Therefore, we provision a per-core register file size of \textit{2*(crossbar dimension)*(\# crossbars per core)}.
This size retains storage density while addressing the buffering requirements in the common case as shown in Section~\ref{sec:results-sweep}.
For window-based computations such as pooling layers that have a large number of intervening instructions (due to non-sequential data access across rows), the compiler spills registers to tile memory (Section~\ref{sec:compiler-reg-alloc}).

\subsubsection{ISA implications}\label{sec:isa-implications}
To accommodate the large register file required to match crossbar size, long operands are used in ISA (\textit{src} and \textit{dest} in Table~\ref{tab:isa_over}), which is another motivation for the wide instruction design.
To accommodate moving data between general purpose registers and \textit{XbarIn}/\textit{XbarOut} registers, a \textit{copy} instruction is included.

\subsection{Memory Unit (MU)}\label{sec:core-mu}

The MU interfaces the core with the tile memory via \textit{load} and \textit{store} instructions.
These instructions can be executed at 16-bit word granularity to support random access as motivated in Section~\ref{sec:workload-access-pattern}.
However, the instructions also take a \textit{vec-width} operand for wide vector loads caused by sequential access patterns.
Vector loads also use temporal SIMD (Section~\ref{sec:core-vfu}) to minimize fetch/decode energy consumption.

\begin{figure}[t]
  \centering
  \includegraphics[width=0.9\columnwidth]{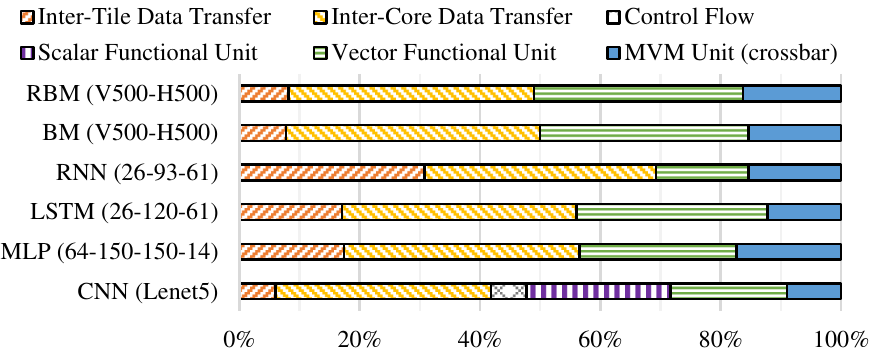}
  \vspace{-0.3cm}
  \caption{Static instruction usage showing the importance of different execution units.}\label{fig:results-instruction-usage}
  \vspace{-0.3cm}
\end{figure}

\subsection{Static Instruction Usage}\label{sec:core-usage}

Figure~\ref{fig:results-instruction-usage} shows the breakdown of the static instruction count for six different ML workloads.
The breakdown demonstrates that MVM alone is insufficient to support all types of workloads, and that the ISA and functional units proposed can be used to bridge that gap.
The ratio of instructions requiring MVMU versus VFU varies depending on the number of matrix versus vector transformation layers in the network.
CNNs additionally use control flow instructions as discussed in Section~\ref{sec:workload-input-compute}.
Deeper (or wider) versions of the same networks tend to have a similar instruction breakdown, except for data movement instructions which tend to be higher to implement larger matrices spanning multiple cores and tiles.

\subsection{Summary}

In summary, the core architecture provides programmability while maintaining crossbar area efficiency.
It features an instruction pipeline exposed by an ISA to support a wide variety of ML workloads.
The use of temporal SIMD and ROM-Embedded RAM enable linear, nonlinear, and transcendental vector operations.
Data movement optimizations are enabled via input shuffling, proper sizing of the register file, and flexible memory access instructions.
\ignore{
Data movement optimizations are enabled in collaboration with the compiler via input shuffling, proper sizing of the register file, and flexible memory access instructions.
}

\section{Tile Architecture}\label{sec:tile}

\begin{figure}[t]
  \centering
  \includegraphics[width=0.9\columnwidth]{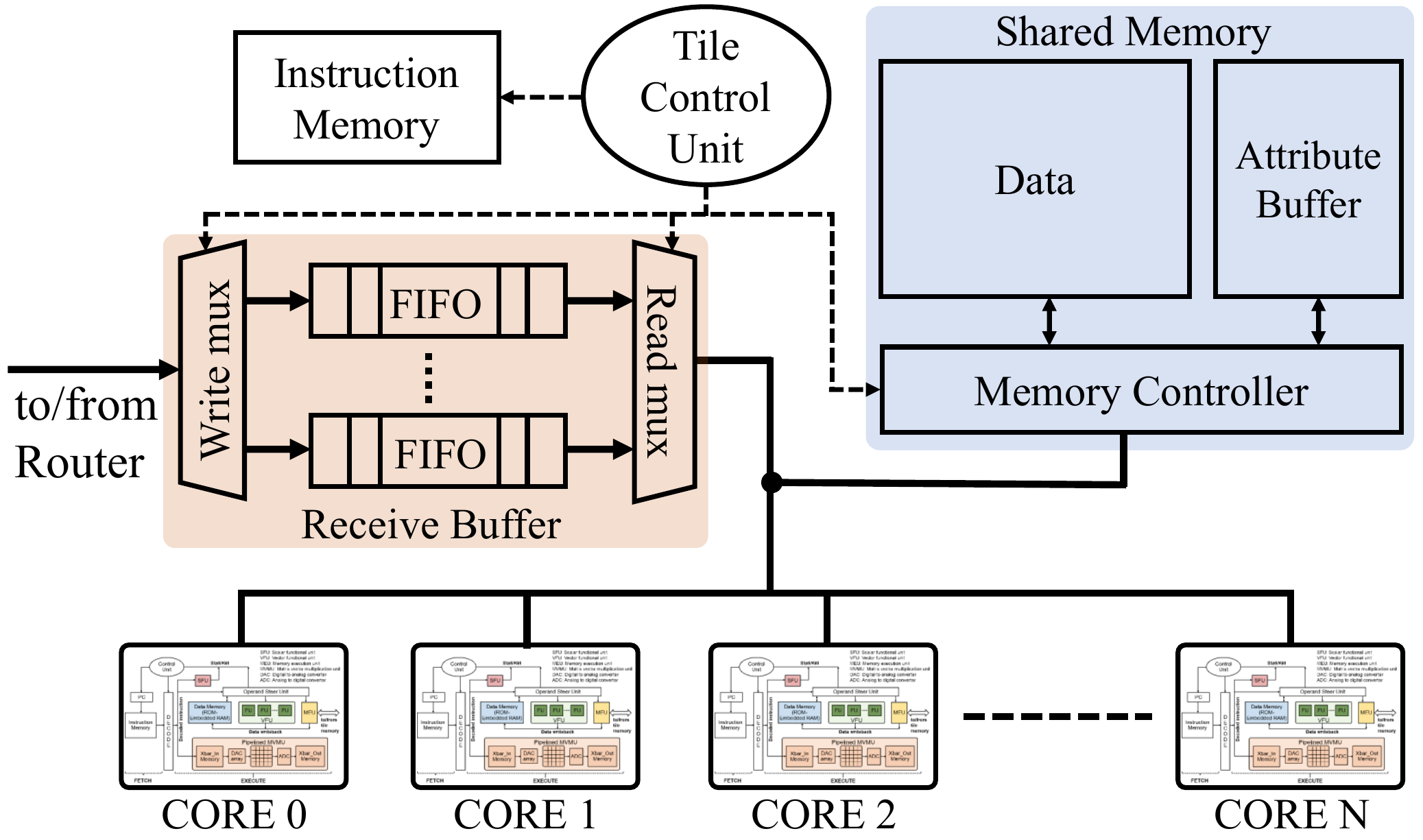}
  \vspace{-0.3cm}
  \caption{Tile Architecture}\label{fig:tile-architecture}
  \vspace{-0.5cm}
\end{figure}

\Fig{fig:tile-architecture} illustrates the architecture of a tile.
A tile is comprised of multiple cores connected to a shared memory.
The tile instruction memory holds \textit{send} and \textit{receive} instructions that move data between tiles.
The shared memory and receive buffer are described in the following subsections.

\subsection{Shared Memory}

The shared memory facilitates communication across cores and tiles.
Our shared memory design follows two key principles: (1) enabling inter-core synchronization, and (2) sizing the shared memory to preserve storage density.

\subsubsection{Inter-core synchronization}\label{sec:intra-tile-comm}

\begin{figure}[t]
  \centering
  \includegraphics[width=0.7\columnwidth]{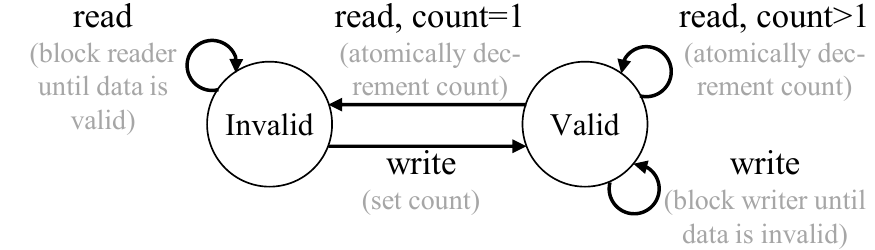}
  \vspace{-0.3cm}
  \caption{Inter-core synchronization mechanism}\label{fig:inter-core-sync}
  \vspace{-0.5cm}
\end{figure}

Synchronization between cores happens when the output of one layer is sent as input to the next layer.
It also happens within a layer if a large weight matrix is partitioned across multiple cores and tiles and partial MVM results need to be aggregated together.
To enable synchronization, we augment the shared memory with an attribute buffer that has two attributes per data entry: valid and count.
The use of valid and count is illustrated in \Fig{fig:inter-core-sync}.
This mechanism enables consumer cores to block until producer cores have written their values, and ensures that producer cores do not overwrite data until it is consumed.

\subsubsection{Sizing the shared memory}\label{sec:tile-shared-memory}

ML workloads may require buffering large number of inputs to utilize their weight reuse pattern.
However, a large shared memory degrades the crossbar storage density.
PUMA's spatial architecture enables programming \textit{inter-core/tile pipelines} that exploit the inter-layer parallelism in ML workloads with weight reuse.
These pipelines can be used to maintain throughput while keeping the shared memory size small.
The pipeline parallelism is based on the observation that we do not require all the outputs of previous layer to start the current layer computation.
For example, LSTMs process a sequence of vectors with $S*N$ inputs per layer, where $S$ is the number of vectors per input sequence and $N$ is vector size.
A layer can begin its computation as soon as its first $N$ inputs are available.
Section ~\ref{sec:result-optimizations} discusses the sizing requirements for different workloads and the impact on energy consumption.

\subsection{Receive Buffer}\label{sec:tile-recv}

The receive buffer is an $N \times M$ structure with $N$ FIFOs, each with $M$ entries.
FIFOs ensure that data being sent from the same source tile is received in the same order.
Having multiple FIFOs enables multiple source tiles to send data concurrently using different FIFOs.
It also enables data to be received through the network independently of receive instruction ordering in the program.
This independence is important because receive instructions are executed in program order in a blocking manner for hardware simplicity.

Each send and receive instruction has a \textit{fifo-id} operand that specifies the receiving FIFO to be used for incoming data.
Using the FIFO ID instead of the sender tile ID provides additional flexibility for the compiler to apply FIFO virtualization, where a FIFO can be used by different sender tiles in different programs or program phases while keeping the number of physical FIFOs small.
The key insight is that a typical ML layer will receive inputs from the tiles mapped to the previous layer only.
Therefore, using 16 FIFOs (despite the node having 138 tiles) supports workloads with up to \textit{(16 tiles)*(8 cores)*(2 MVMU)*128} previous layer activations, which suffices for large-scale ML workloads.

\subsection{Summary}

In summary, PUMA tiles enable inter-core synchronization, inter-core/tile pipelines to contain shared memory size, and FIFO virtualization for efficient inter-tile communication.

\section{Compiler}\label{sec:compiler}

PUMA is a spatial architecture (not a data-parallel processor) which means that each core/tile executes a different set of instructions.
Writing different code for each core/tile is not scalable as applications grow in size and complexity.
A compiler is therefore mandatory for programmer productivity.
This section describes key aspects of the compiler, while other implementation details of the compiler and the rest of the software stack are described in another paper~\cite{ambrosi2018hardware}.

\subsection{Programming Interface}

\begin{figure}[t]
  \centering
  \includegraphics[width=0.95\columnwidth]{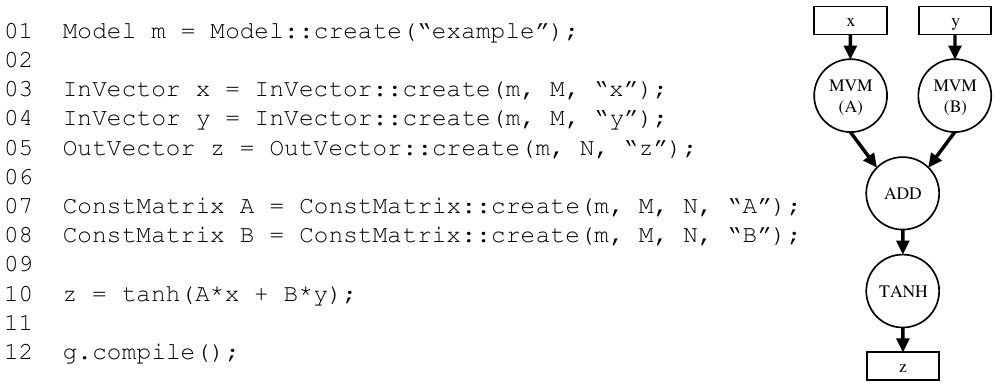}
  \vspace{-0.3cm}
  \caption{Simple Code Example}\label{fig:code-example}
  \vspace{-0.3cm}
\end{figure}

The PUMA compiler is a runtime compiler implemented as a C++ library.
A simple code example is shown in \Fig{fig:code-example}.
The programmer first creates a model (line 01) with input/output vectors (lines 03-05) and constant matrices (lines 07-08).
The programmer may also create vector streams which are useful for setting up inter-core/tile pipelines (see Section~\ref{sec:tile-shared-memory}).
The programmer then describes a computation (line 10) which executes at run time to build a graph of the model (\Fig{fig:code-example} on the right).
Finally, the model is compiled (line 12) to generate PUMA assembly code for each core and tile.
In addition to this native interface, ONNX bindings are also provided for further adoption and interoperability, enabling the compilation of models written in popular DNN frameworks such as Caffe2, PyTorch, Cognitive Toolkit, MXNet, and others.

\subsection{Graph Partitioning}

\begin{figure}[t]
  \centering
  \includegraphics[width=0.9\columnwidth]{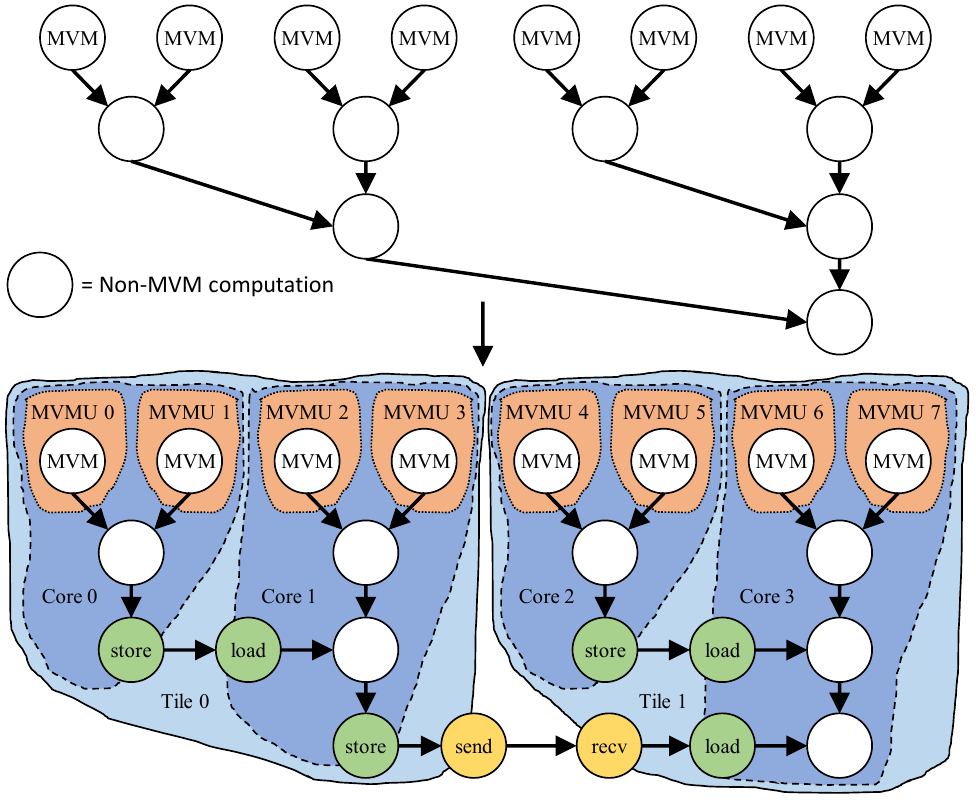}
  \vspace{-0.3cm}
  \caption{Graph Partitioning Example}\label{fig:partition-data}
  \vspace{-0.3cm}
\end{figure}

The first step in the compilation process is graph partitioning.
The compiler divides tensors into 2D tiles, each the size of one MVMU, with appropriate padding, and divides the corresponding vectors and operations in the model accordingly.
Next, the graph is hierarchically partitioned, distributing sub-graphs to different MVMUs, cores, and tiles as shown in the example in \Fig{fig:partition-data}.
The partitioning scheme used in this paper prioritizes placing MVMUs that feed to the same outputs together on the same core/tile, followed by those that read the same inputs, followed by those that feed each other (i.e., producer-consumer MVMUs).
After partitioning the graph, the compiler inserts load/store operations across cores and allocates shared memory accordingly, reusing memory locations when there is pipelining.
The compiler also inserts send/receive operations across tiles and assigns FIFO IDs accordingly (see Section~\ref{sec:tile-recv}), thereby virtualizing the FIFOs and ensuring there are no conflicts.

\subsection{Instruction Scheduling}

\begin{figure}[t]
  \centering
  \includegraphics[width=0.85\columnwidth]{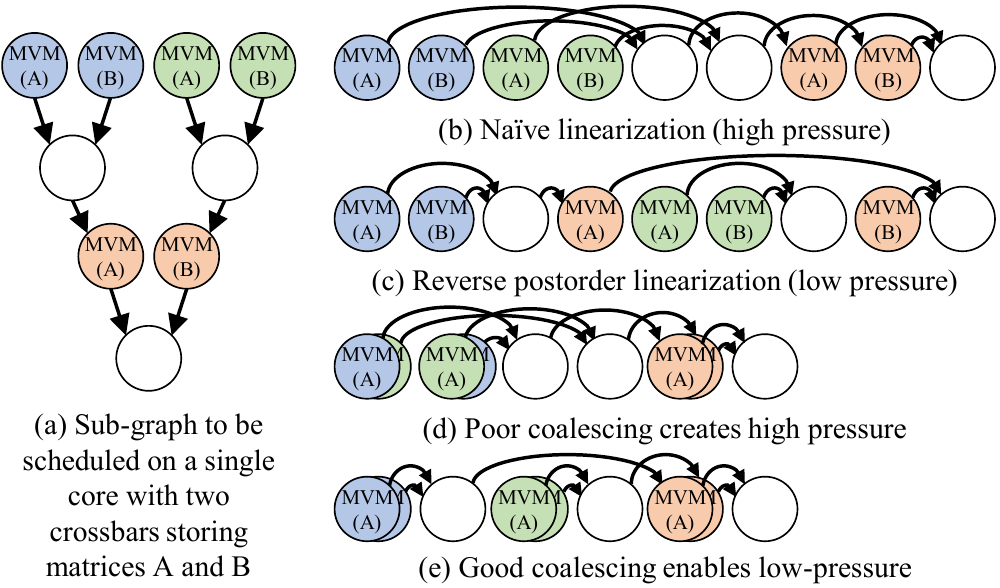}
  \vspace{-0.3cm}
  \caption{Instruction Scheduling Example}\label{fig:instruction-scheduling}
  \vspace{-0.3cm}
\end{figure}

After the graph is partitioned into a sub-graph for each core/tile, the compiler schedules instructions by linearizing each sub-graph.
Instruction scheduling has three main objectives: reducing register pressure, capturing ILP of MVM operations, and avoiding deadlock.

\subsubsection{Reducing register pressure}\label{sec:compiler-instruction-pressure}

There are many possible ways to linearize the sub-graph, as long as source operations are visited before their destinations to enforce data dependencies.
We use a reverse post-order traversal which prioritizes consuming produced values before producing new ones.
This ordering reduces the number of live values, hence the register pressure.
\Fig{fig:instruction-scheduling}(b) and (c) show two correct linearizations of the sub-graph in \Fig{fig:instruction-scheduling}(a).
Reverse postorder in \Fig{fig:instruction-scheduling}(c) results in fewer live values at any point in time.

\subsubsection{Capturing ILP of MVM operations}\label{sec:compiler-coalescing}

As explained in Section~\ref{sec:core-mvmu-coalesce}, it is desirable to run different MVMUs in the same core simultaneously.
The compiler must therefore fuse independent MVM operations on different MVMUs in the same core.
We call this optimization \textit{MVM coalescing}.
MVMs can be coalesced when there are no data dependences between them.
It is also desirable to coalesce MVMs whose results are consumed soon after one another to reduce register pressure.
The compiler's strategy for coalescing is to first look for coalescing candidates among MVMs that are tiles of the same large MVM operation.
Once those are exhausted, the remaining MVMs are coalesced by traversing the graph in reverse poster-order (before linearization) and fusing each MVM node with the first eligible candidates in the traversal order.
The dependence information is updated every time a fusion takes place.
Finally, the graph is traversed one last time to perform the linearization.
\Fig{fig:instruction-scheduling}(d) and (e) show two example linearizations, with \Fig{fig:instruction-scheduling}(e) following the proposed approach resulting in fewer live values at once.

\begin{figure}[t]
  \centering
  \includegraphics[width=0.9\columnwidth]{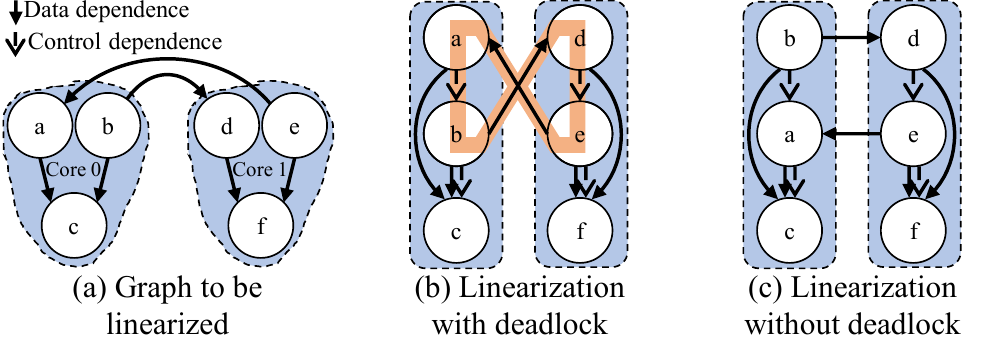}
  \vspace{-0.3cm}
  \caption{Deadlock Avoidance Example}\label{fig:deadlock}
  \vspace{-0.3cm}
\end{figure}

\subsubsection{Avoiding deadlock}

Linearizing the sub-graph of each core introduces control edges to the graph.
Since communication across cores is blocking (see Section~\ref{sec:intra-tile-comm}), cycles introduced by improper linearization cause deadlock as shown in the example in \Fig{fig:deadlock}(b).
For this reason, sub-graphs are not linearized independently.
The entire graph is linearized at once placing instruction in the corresponding core/tile sequence to ensure a globally consistent linearization order.

\subsection{Register Allocation}\label{sec:compiler-reg-alloc}

The final step in the compilation is register allocation.
Recall that a core has three sets of registers: \textit{XbarIn}, \textit{XbarOut}, and general purpose.
XbarIn (XbarOut) registers can be written (read) by any non-MVM instruction but can only be read (written) by MVM instructions.
General purpose registers can be read and written by any non-MVM instructions.
Liveness analysis is performed on each of these sets of registers separately.
Conflicting live ranges in \textit{XbarIn} and \textit{XbarOut} registers result in spilling to general purpose registers via \textit{copy} instructions.
Conflicting live ranges in general purpose registers result in spilling to shared memory via load/store instructions.

\subsection{Summary}

In summary, the PUMA compiler provides a high-level programming interface and performs graph partitioning, instruction scheduling, and register allocation to generate low-level assembly.
Instruction scheduling aims at reducing register pressure, MVM colescing, and avoiding deadlock.

\section{Evaluation Methodology}\label{sec:methodology}

\begin{table}[t]
\centering
    \caption{PUMA Hardware Characteristics
    }\label{tab:hardwaretab}
    \vspace{-0.4cm}
    \centering
    \resizebox{1.0\columnwidth}{!}{
        \begin{tabular}{|l|c|c|c|c|}
    \hline
    \multicolumn{5}{|c|}{\textbf{PUMA Tile at 1GHz on 32nm Technology node}}\\
    \hline
    \textbf{Component}      & \textbf{Power (mW)}   & \textbf{Area (mm\textsuperscript{2})} & \textbf{Parameter}     & \textbf{Specification} \\
    \hline
    Control Pipeline        & 0.25                  & 0.0033                                & \# stages              & 3                      \\
    \hline
    Instruction Memory      & 1.52                  & 0.0031                                & capacity               & 4KB                    \\
    \hline
    Register File           & 0.477                 & 0.00192                               & capacity               & 1KB                    \\
    \hline
    MVMU                    & 19.09                 & 0.012                                 & \# per core            & 2                      \\
                            &                       &                                       & dimensions             & 128$\times$128 \\
    \hline
    VFU                     & 1.90                  & 0.004                                 & width                  & 1                      \\
    \hline
    SFU                     & 0.055                 & 0.0006                                & -                      & -                      \\
    \hline
    Core                    & 42.37                 & 0.036                                 & \# per tile            & 8                      \\
    \hline
    Tile Control Unit       & 0.5                   & 0.00145                               & -                      & -                      \\
    \hline
    Tile Instruction Memory & 1.91                  & 0.0054                                & capacity               & 8KB                    \\
    \hline
    Tile Data Memory        & 17.66                 & 0.086                                 & capacity               & 64KB                   \\
                            &                       &                                       & technology             & eDRAM                  \\
    \hline
    Tile Memory Bus         & 7                     & 0.090                                 & width                  & 384 bits               \\
    \hline
    Tile Attribute Memory   & 2.77                  & 0.012                                 & \# entries             & 32K                    \\
                            &                       &                                       & technology             & eDRAM                  \\
    \hline
    Tile Receive Buffer     & 9.14                  & 0.0044                                & \# fifos               & 16                     \\
                            &                       &                                       & fifo depth             & 2                      \\
    \hline
    Tile                    & 373.8                 & 0.479                                 & \# per node            & 138                    \\
    \hline
    On-chip Network         & 570.63                & 1.622                                & flit\_size             & 32                     \\
                            &                       &                                      & \# ports               & 4                      \\
                            &                       &                                      & conc                   & 4                      \\
    \hline
    Node                    & 62.5K                 & 90.638                                &  -                     & -                      \\
    \hline
    Off-chip Network        & 10.4K                 & 22.88                                 & type                   & HyperTransport                 \\
    (per node)                        &                       &                                       & link bandwidth                & 6.4 GB/sec              \\
    \hline
\end{tabular}

    }
    \vspace{-0.4cm}
\end{table}

\begin{table}[t]
    \caption{Benchmarking Platforms}\label{tab:sysinfo}
    \vspace{-0.4cm}
    \centering
    \resizebox{0.95\columnwidth}{!}{
        \begin{tabular}{|l|c|} 
         \hline
         \textbf{Name} & \textbf{Platform Characteristics} \\
         \hline
         Haswell & Intel Xeon E5-2650v3, 10-cores per socket, Dual Socket, 128GB DDR4 \\
         \hline
         Skylake & Intel Xeon 8180, 28-cores per socket, Dual Socket, 64GB DDR4\\
         \hline
         Kepler & Nvidia Tesla K80, 2496 CUDA Cores, Dual GPUs (only 1 used), 12GB GDDR5\\
         \hline
         Maxwell & Nvidia Geforce TitanX, 3072 CUDA Cores, 12GB GDDR5 \\
         \hline
         Pascal & Nvidia Tesla P100, 3584 CUDA Cores, 16GB HBM2 \\
        \hline
        \end{tabular}
    }
    \vspace{-0.4cm}
\end{table}

\subsection{PUMAsim}

We have implemented a detailed architectural simulator (PUMAsim) to evaluate the performance and energy consumption of PUMA.
PUMAsim runs applications compiled to the PUMA ISA and provides detailed traces of execution.
The simulator incorporates functionality, timing, and power models of all system components.
The datapath for the PUMA core and tile was designed at Register-Transfer Level (RTL) in Verilog HDL and mapped to IBM 45nm SOI technology using Synopsys Design Compiler which was used to measure the area and power consumption.
Subsequently, these area and power numbers were added to PUMAsim for system-level evaluations of workloads. 
For fair comparison with other application-specific accelerators, the datapath energy numbers have been scaled to the 32nm technology node.
Memory modules are modelled in Cacti 6.0~\cite{Muralimanohar2007CACTI6A} to estimate the area, power, and latency.
The on-chip-network is modelled using the cycle-level Booksim~2.0 interconnection network simulator~\cite{Jiang2013} and Orion~3.0~\cite{Kahng2012} for energy and area models.
We use a chip-to-chip interconnect model similar to DaDianNao's~\cite{chen2014dadiannao} which has also been adopted by other accelerators.
The MVMU power and area models are adapted from ISAAC~\cite{shafiee2016isaac}.
The memristors have a resistance range of $100k\si{\ohm}-1M\si{\ohm}$ and read voltage of 0.5V.
The ADC is based on the Successive Approximation Register (SAR) design, and its area and power were obtained from the ADC survey and analysis~\cite{wang2016neuromorphic, murmann2011adc}.

In the present study, we do not compromise ML accuracy as we conservatively choose 2-bit memristor crossbar cells.
Note that laboratory demonstrations have shown up to 6-bit capabilities~\cite{Hu2018MNIST}.
We use 16 bit fixed-point precision that provides very high accuracy in inference applications~\cite{chen2014dadiannao, shafiee2016isaac}.
Table~\ref{tab:hardwaretab} shows the PUMA configuration used in our analysis and lists the area-energy breakdown of PUMA components.

\begin{table*}[t]
\centering
    \caption{Benchmarks}\label{tab:benchmarkchar}
    \vspace{-0.4cm}
    \centering
    \resizebox{0.75\textwidth}{!}{
        \begin{tabular}{|c|c|c|c|c|c|c|c|c|}
    \hline
     \textbf{DNN Type} & \textbf{Application} & \textbf{DNN Name} & \textbf{\# FC Layers} & \textbf{\# LSTM Layers} & \textbf{\# Conv Layers}  & \textbf{\# Parameters}  & \textbf{Non-linear Function} & \textbf{Sequence Size} \\
    \hline
    \hline
    MLP & Object & MLP\textsubscript{L4} & 4 & - & -& 5M & Sigmoid & - \\
    & Detection & MLP\textsubscript{L5} & 5 & - & - & 21M & Sigmoid & - \\
    \hline
    Deep & Neural Machine & NMT\textsubscript{L3} & 1 & 6 (3 Enc.,3 Dec., 1024 cells) & - & 91M & Signmod, Tanh & 50 \\
    LSTM & Translation & NMT\textsubscript{L5} & 1 & 10 (5 Enc.,5 Dec., 1024 cells) & - & 125M & Signmod, Tanh & 50 \\
    \hline
    Wide & Language & BigLSTM & 1 & 2 (8192 cell, 1024 proj) & - & 856M & Signmod, Tanh, LogSoftMax & 50 \\
    LSTM & Modelling & LSTM-2048 & 1 & 1 (8192 cell, 2048 proj) & - & 554M & Sigmoid, Tanh, LogSoftMax & 50\\
    \hline
    CNN & Image & Vgg16 &3&- &13 &136M & ReLU & - \\
     & Recognition & Vgg19 & 3 & - & 16 & 141M& ReLU & -\\
     \hline
\end{tabular}
    }
    \vspace{-0.2cm}
\end{table*}

\begin{figure*}[t]
  \centering
  \includegraphics[width=0.85\textwidth]{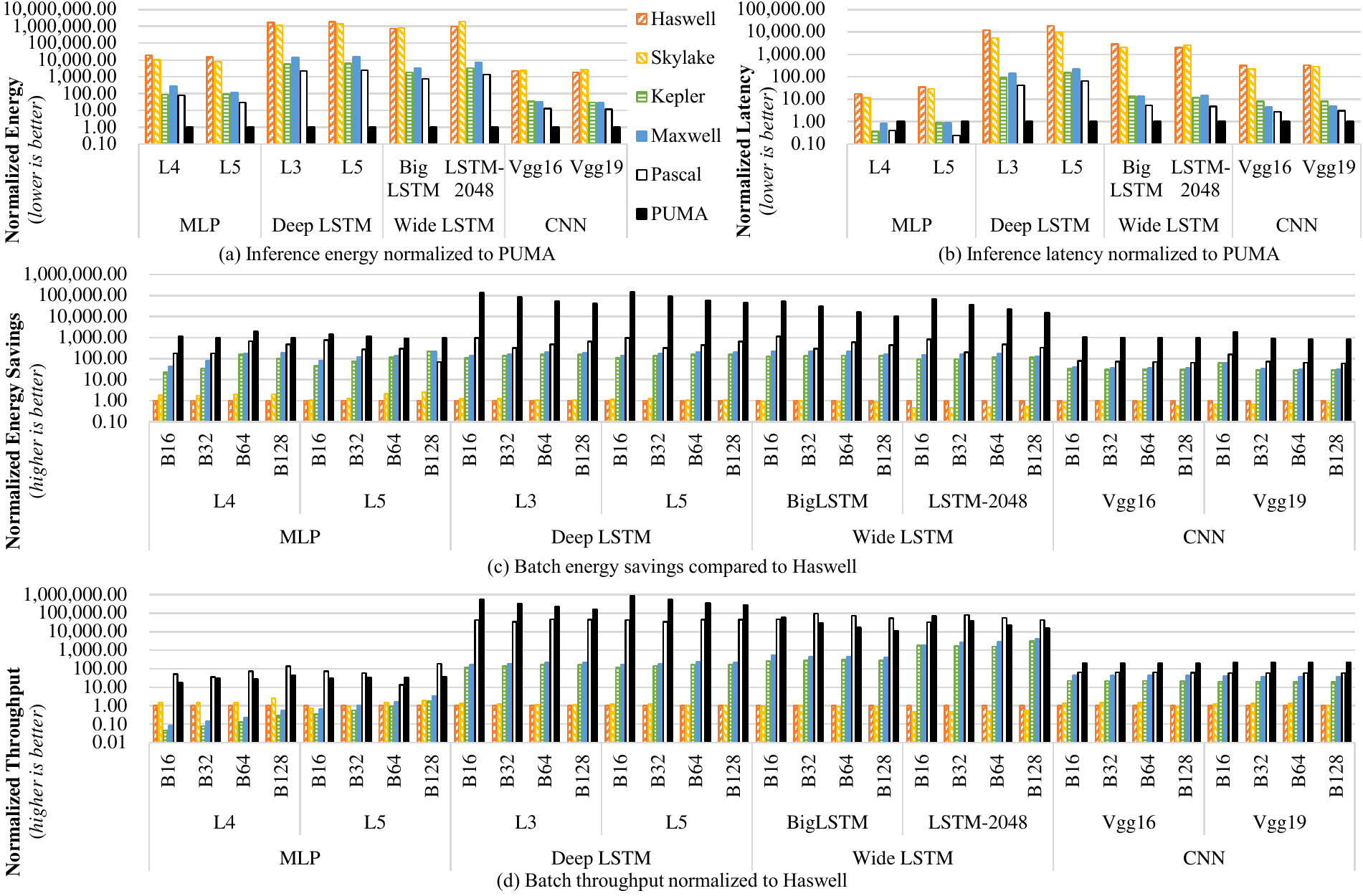}
    \vspace{-0.3cm}
  \caption{Energy and Latency Results}\label{fig:results-latency-energy}
    \vspace{-0.4cm}
\end{figure*}

\subsection{System and Workloads}

We choose popular server grade CPUs and GPUs (listed in Table~\ref{tab:sysinfo}), Google TPU~\cite{Jouppi:2017:IPA:3079856.3080246} (CMOS-based ASIC) and ISAAC~\cite{shafiee2016isaac} (application specific memristor-based accelerator) for evaluating PUMA.
To measure power consumption of CPUs and GPUs, we used management tools such as board management control (BMC) and nvidia-smi respectively.
For GPUs, we do not include full system power, just the board/device power.
We run multiple iterations of the benchmarks on the GPU, discarding the longer warmup iterations and reporting results from the faster and more stable iterations.

Torch7~\cite{collobert2011torch7} was used to execute the ML models for CPUs and GPUs.
The PUMA compiler was used to compile models for PUMA.
These models used are listed in Table~\ref{tab:benchmarkchar}.

\ignore{
\begin{figure*}[t]
  \centering
  \includegraphics[width=0.85\textwidth]{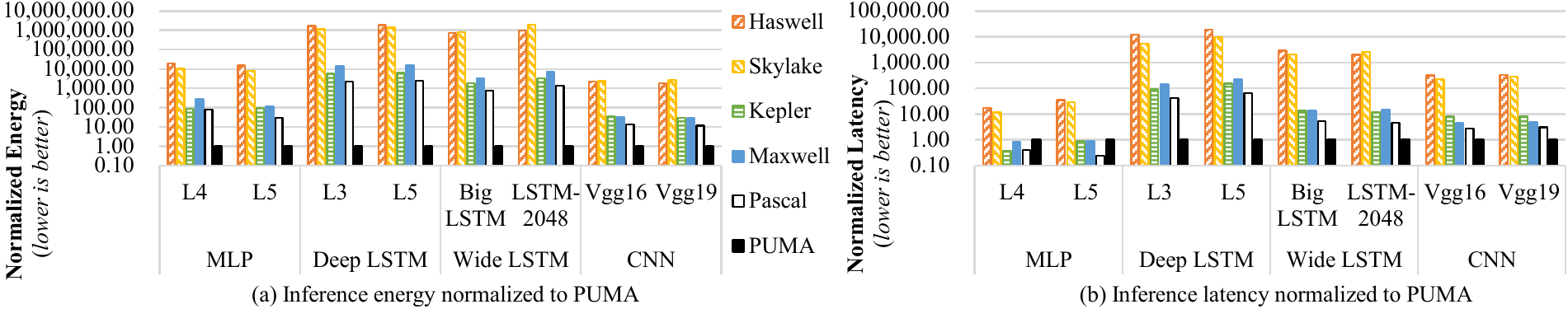}
  \shrinkBeforeFigureCaption
  \caption{Inference Energy and Latency Results}\label{fig:results-inference-latency-energy}
  \shrinkAfterFigure
\end{figure*}

\begin{figure*}[t]
 \centering
 \includegraphics[width=0.85\textwidth]{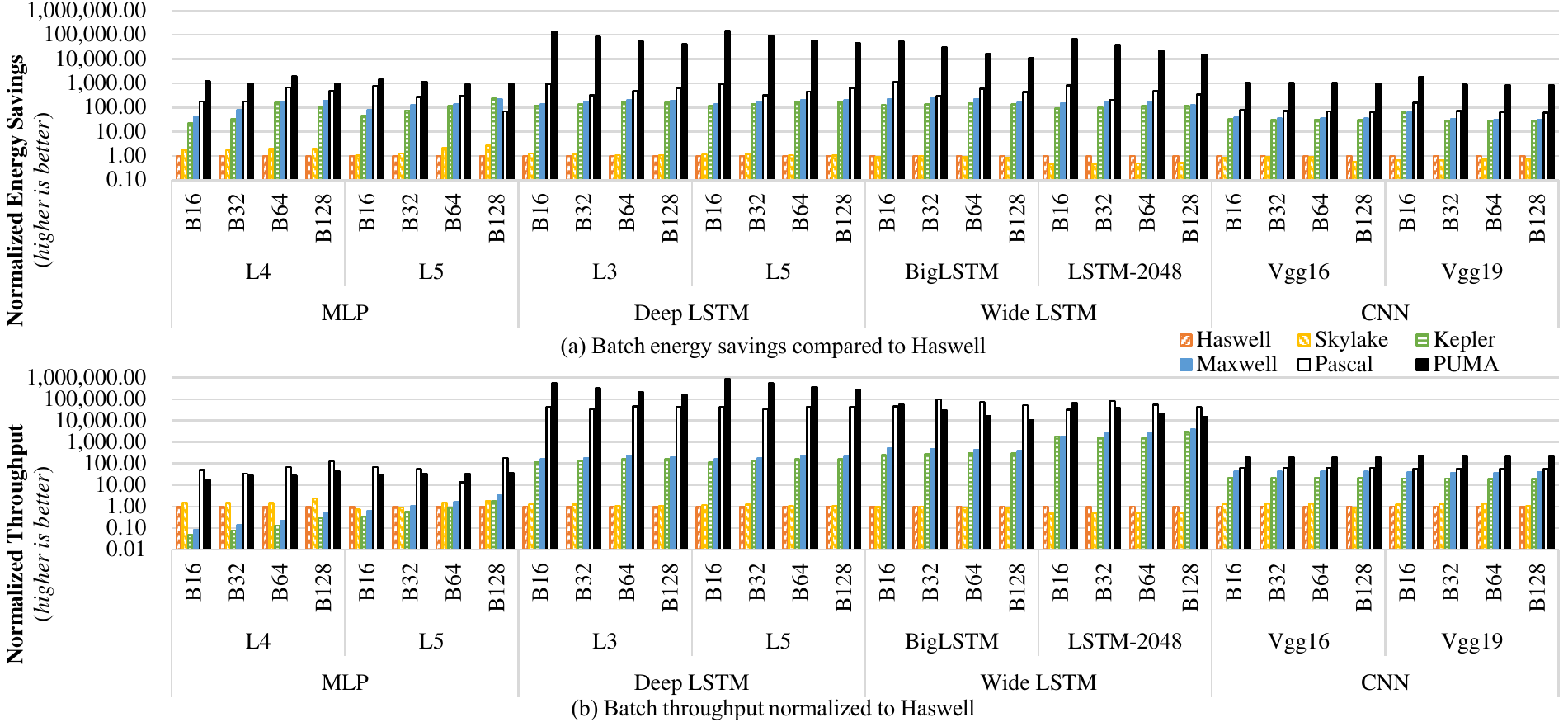}
 \shrinkBeforeFigureCaption
  \caption{Batch Inference Throughput and Energy Results}\label{fig:results-batch-throughput-energy}
  \shrinkAfterFigure
\end{figure*}
}    \vspace{-0.5cm}

\section{Results}\label{sec:results}

\subsection{Inference Energy}\label{sec:inference-energy}

\Fig{fig:results-latency-energy}(a) shows PUMA inference energy compared to other platforms.
PUMA achieves massive energy reduction across all benchmarks for all platforms.
Energy improvements come from two sources: lower MVM compute energy from crossbars and lower data movement energy by avoiding weight data movement.

CNNs show the least energy reduction over CMOS architectures (11.7$\times$-13.0$\times$ over Pascal).
Recall that CNNs have a lot of weight reuse because of the sliding window computation (discussed in Section~\ref{sec:workload-input-compute}).
Hence, CMOS architectures can amortize DRAM accesses of weights across multiple computations.
For this reason, PUMA's energy savings in CNNs come primarily from the use of crossbars for energy efficient MVM computation.

MLPs and LSTMs have little or no weight reuse (discussed in Section~\ref{sec:workload}). 
Therefore, in addition to efficient MVM computation, PUMA has the added advantage of eliminating weight data movement.
For this reason, we see much better energy reductions for MLPs (30.2$\times$-80.1$\times$ over Pascal), Deep LSTMs (2,302$\times$-2,446$\times$ over Pascal), and Wide LSTMs (758$\times$-1336$\times$ over Pascal).

LSTMs (both deep and wide) show better reductions than MLPs because they have much larger model sizes (see \# Parameters in Table~\ref{tab:benchmarkchar}).
As model grows in size, weight data grows at $O(n^2)$ and input data grows at $O(n)$.
For this reason, we see an increasing disparity between CMOS architectures which move both weight and input data, and PUMA which only moves input data.

Wide LSTMs have few layers (1-2) with very large matrices, whereas Deep LSTMs have many layers (6-10) with smaller matrices.
The large matrices in Wide LSTMs span mutiple PUMA cores/tiles to compute one logical MVM, incurring higher intra-layer data movement overheads.
Hence, Deep LSTMs show higher energy benefits than Wide LSTMs.

\subsection{Inference Latency}\label{sec:inference-latency}

\Fig{fig:results-latency-energy}(b) shows PUMA inference latency compared to other evaluated platforms.
PUMA achieves latency improvements across all platforms except MLPs on some GPUs.
Latency improvements come from three sources: lower MVM compute latency from crossbars, no weight data access latency, and spatial architecture pipelining which exploits inter-layer parallelism.

CNNs show the least latency improvement over CMOS architectures (2.73$\times$-2.99$\times$ over Pascal).
Since CNNs are compute-bound, CMOS architectures can hide the data access latency.
Thus, PUMA's primary latency improvements in CNNs come from the use of crossbars for low-latency MVM computation and spatial architecture pipelining.

LSTMs on the other hand are memory-bound.
PUMA has the added advantage of eliminating weight data access latency in addition to low-latency MVM computation.
For this reason, we see much better latency improvements for Deep LSTMs (41.6$\times$-66.0$\times$ over Pascal) and Wide LSTMs (4.70$\times$-5.24$\times$ over Pascal) than we see for CNNs.
In comparing Deep and Wide LSTMs, Deep LSTMs have more layers than Wide LSTMs, hence more inter-layer parallelism to exploit spatial architecture pipelining (see \#LSTM Layers in Table~\ref{tab:benchmarkchar}).
Moreover, Deep LSTMs have less intra-layer communication than Wide LSTMs, hence lower data access latency.

MLPs show slowdown compared to some GPU datapoints (0.24$\times$-0.40$\times$ compared to Pascal).
The reason is that despite MLPs being memory-bound, the sizes of MLPs are typically small enough.
Hence, the memory bandwidth bottleneck is not as pronounced, so they perform fairly well on GPUs.
Moreover, MLPs have no inter-layer parallelism so they do not exploit spatial architecture pipelining (Section~\ref{sec:tile-shared-memory}).
Nevertheless, PUMA's order of magnitude energy reduction is still beneficial for MLPs in energy-constrained environments.

\subsection{Batch Throughput and Energy}

Inference applications are not usually intended for large batch sizes due to real-time application requirements.
Nevertheless, CMOS architectures perform well with large batch sizes because of the weight reuse that data-batching exposes.
For this reason, we compare PUMA's batch energy and throughput with the other platforms in \Fig{fig:results-latency-energy}(c) and (d) respectively.
Batch sizes of 16, 32, 64, and 128 are used.

PUMA continues to have superior energy efficiency across all benchmarks and batch sizes.
It also delivers better throughput in most cases, except when compared to Pascal on MLPs and Wide LSTMs.
The benefits slightly decrease with larger batches because they expose more weight reuse which benefits CMOS architectures while PUMA's efficiency remains constant across batch sizes.
Nevertheless, PUMA continues to perform well even when the batch size is very large.

\subsection{Comparison with ML Accelerators}
\subsubsection{Google TPU}
Table~\ref{tab:result-compare} compares key technology features and efficiency metrics for TPU~\cite{Jouppi:2017:IPA:3079856.3080246} and PUMA.
PUMA has 8.3$\times$ higher peak area-efficiency (TOPS/s/mm\textsuperscript{2}) than TPU.
Since PUMA does not rely on weight reuse to improve throughput like CMOS architectures do, its area-efficiency remains constant across workloads and batch sizes.
On the other hand, TPU's peak throughput is almost an order of magnitude lower for applications with low data reuse due to its inability to amortize weight data movement.
PUMA has 64.4$\times$, 193$\times$, and 9.7$\times$ higher area-efficiency than TPU for MLP, LSTM, and CNN respectively for the best TPU batch size.

\begin{table}[t]
\centering
    \caption{Comparison with ML Accelerators}\label{tab:result-compare}
    \vspace{-0.4cm}
    \centering
    \resizebox{0.98\columnwidth}{!}{
        \begin{tabular}{|l|c|c|c|}
    \hline
    \textbf{Platform}                           & \textbf{PUMA}         & \textbf{TPU}~\cite{Jouppi:2017:IPA:3079856.3080246}     & \textbf{ISAAC}~\cite{shafiee2016isaac}  \\
    \hline
    \hline
    \textbf{Year}                               & 2018                  & 2017             & 2016            \\
    \hline
    \textbf{Technology}                         & CMOS(32nm)-Memristive & CMOS(28nm)       & CMOS(32nm)-Memristive  \\
    \hline
    \textbf{Clock (MHz)}                        & 1000                  & 700              & 1200       \\
    \hline
    \textbf{Precision}                          & 16-bit fixed point    & 16-bit fixed point    & 16-bit fixed point    \\
    \hline
    \textbf{Area (\textit{mm$^2$)}}             & 90.6                  & 330*             & 85.4       \\
    \hline
    \textbf{Power (W)}                          & 62.5                  & 45               & 65.8       \\
    \hline
    \hline
    \textbf{Peak Throughput (TOPS/s$^\dagger$)} & 52.31                 & 23$^\ddagger$    & 69.53      \\
    \hline
    \textbf{Peak AE (TOPS/s/mm$^2$)}            & \textbf{0.58}         & \textbf{0.07}    & \textbf{0.82}       \\
    \hline
    \textbf{Peak PE (TOPS/s/W)}                 &  0.84                 & 0.51             & 1.06       \\
    \hline
    \hline
    \textbf{Best AE - MLP}                      & 0.58                  & 0.009            & -     \\
    \hline
    \textbf{Best AE - LSTM}                     & 0.58                  & 0.003            & -     \\
    \hline
    \textbf{Best AE - CNN}                      & 0.58                  & 0.06             & 0.82      \\
    \hline
    \hline
    \textbf{Best PE - MLP}                      & 0.84                  & 0.07             & -     \\
    \hline
    \textbf{Best PE - LSTM}                     & 0.84                  & 0.02             & -     \\
    \hline
    \textbf{Best PE - CNN}                      & 0.84                  & 0.48             & 1.06      \\
    \hline
    \multicolumn{4}{l}{* Less than or equal to half of Haswell's die area~\cite{Jouppi:2017:IPA:3079856.3080246}} \\
    \multicolumn{4}{l}{$^\dagger$ Tera operations per second (multiply and add are counted as two separate operations)} \\
    \multicolumn{4}{l}{$^\ddagger$ 92 TOPS for 8-bit arithmetic, scaled by 4 for 16-bit arithmetic~\cite{Jouppi:2017:IPA:3079856.3080246}} \\
\end{tabular}
    }
    \vspace{-0.4cm}
\end{table}

\begin{table}[t]
\centering
    \caption{Programmability Comparison with ISAAC}\label{tab:puma-issac-table}
    \vspace{-0.4cm}
    \centering
    \resizebox{0.98\columnwidth}{!}{
        
\begin{tabular}{|l|c|c|}
\hline
\textbf{Platforms}                     & \textbf{PUMA}                                                                                                            & \textbf{ISAAC}                               \\ \hline
\multirow{2}{*}{\textbf{Architecture}} & \begin{tabular}[c]{@{}c@{}}Instruction execution pipeline,\\ flexible inter-core synchronization\end{tabular}            & Application specific state machine           \\ \cline{2-3} 
                                       & Vector Functional Unit, ROM-Embedded RAM                                                                                 & Sigmoid unit                                 \\ \hline
\textbf{Programmability}               & Compiler-generated instructions (per tile \& core)                                                                       & Manually configured state machine (per tile) \\ \hline
\textbf{Workloads}                     & \begin{tabular}[c]{@{}c@{}}CNN, MLP, LSTM, RNN, GAN, BM, RBM,\\ SVM, Linear Regression, Logistic Regression\end{tabular} & CNN                                          \\ \hline
\end{tabular}
    }
    \vspace{-0.4cm}
\end{table}

PUMA has 1.65$\times$ higher peak power-efficiency (TOPS/s/W) than TPU, with similar trends and reasoning as area-efficiency for specific workloads.
We expect PUMA's power-efficiency advantage over TPU to grow by over 3$\times$, as silicon processes scale from 32nm to 7nm and 5nm.
Thanks to PUMA's higher peak throughput, we can follow the power reduction scaling curve at constant performance.
Conversely, to narrow the performance gap, TPU would follow a scaling curve closer to the performance increase curve at constant power.
Note that the former scaling curve is much faster (up to $\sim$40\% power reduction per silicon node compared with $\sim$20\% performance increase).
Further, ADC power efficiency has also been following similar and very fast scaling trend with $\sim$2$\times$ power reduction per 1.8 years at the same sampling rate~\cite{murmann2015race}.
\ignore{
Consequently, we project that our power-efficiency advantage over TPU will increase by over 3$\times$ after scaling PUMA and TPU to 7nm.
}

\begin{figure*}[t]
  \centering
  \includegraphics[width=0.85\textwidth]{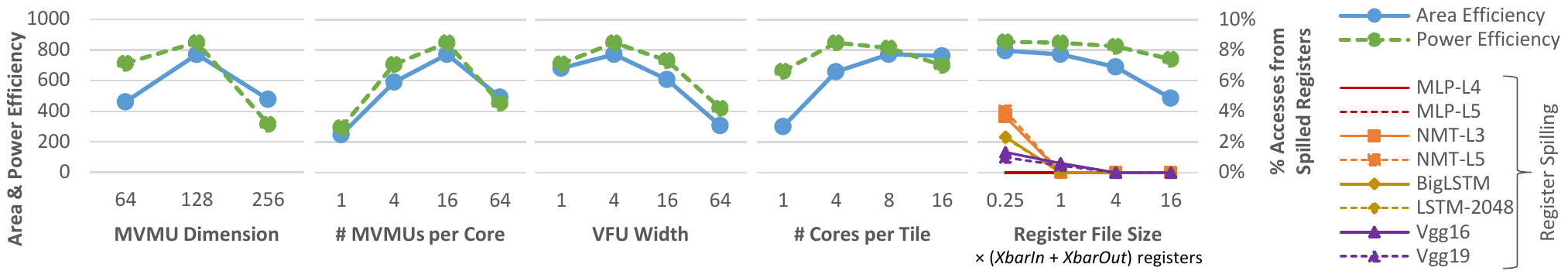}
  \vspace{-0.3cm}
  \caption{Design Space Exploration: Tile Area Efficiency in GOPS/s/mm\textsuperscript{2} and Tile Power Efficiency in GOPS/s/W}\label{fig:design-space}
  \vspace{-0.3cm}
\end{figure*}

\subsubsection{ISAAC}

Table~\ref{tab:result-compare} compares the peak area and power efficiency of PUMA with ISAAC~\cite{shafiee2016isaac}, a memristor-based accelerator customized for CNNs.
PUMA has 20.7\% lower power efficiency and 29.2\% lower area efficiency than ISAAC due to the overhead of programmability.
Table~\ref{tab:puma-issac-table} compares the programmability of PUMA and ISAAC.

\subsubsection{PUMA with Digital MVMUs}

To demonstrate the importance of analog computing for MVMU efficiency, we compare PUMA with a hypothetical equivalent that uses digital MVMUs.
A memristive 128$\times$128 MVMU performs 16,384~MACs in 2304~ns consuming 43.97~nJ.
A digital MVMU would require 8.97$\times$ more area to achieve the same latency and would consume 4.17$\times$ more energy.
Using a digital MVMU would increase the total chip area of the accelerator by 4.93$\times$ for the same performance and would consume 6.76$\times$ energy (factoring in data movement energy due to increased area).

\subsubsection{Tensor Cores}

Nvidia V100 GPUs with tensor cores (FP16) can be up to $6\times$ more energy-efficient (architecture, tensor cores, and half-precision) than Pascal GPUs.
Therefore, PUMA can still achieve energy gains over GPUs with tensor cores, despite the technology difference (PUMA: 32nm, V100: 12nm). 

\begin{table}[t]
\centering
    \caption{Evaluation of Optimizations}\label{tab:optimizations}
    \vspace{-0.4cm}
    \centering
    \resizebox{\columnwidth}{!}{
        \begin{tabular}{|l|c|c|c|c|c|}
    %\multicolumn{6}{c}{Register pressure column shows \% accesses from spilled registers}                                                                          \\
    %\multicolumn{6}{c}{MVM coalescing column shows latency reduction caused by optimization (lower is better).}                                                          \\
    %\multicolumn{6}{c}{Remaining columns show energy reduction caused by optimization (lower is better).}                                                          \\
    \hline
    \multirow{4}{*}{\textbf{Workload}}      & \textbf{Input}     & \textbf{Shared}          & \textbf{Graph}            & \textbf{Register}   & \textbf{MVM}          \\
                                            & \textbf{Shuffling} & \textbf{Memory Sizing}   & \textbf{Partitioning}     & \textbf{Pressure}   & \textbf{Coalescing}   \\
                                            & (energy reduction, & (energy reduction,       & (energy reduction,        & (\% accesses from  & (latency reduction,     \\
                                            & lower is better)   & lower is better)         & lower is better)          & spilled registers) & lower is better)     \\
    \hline
    MLP\textsubscript{L4}                                & -                  & 0.70$\times$             & 0.81$\times$             & 0\%               & 0.60$\times$          \\
    \hline
    MLP\textsubscript{L5}                                & -                  & 0.66$\times$             & 0.79$\times$             & 0\%               & 0.66$\times$          \\
    \hline
    NMT\textsubscript{L3}                                & -                  & 0.65$\times$             & 0.65$\times$             & 0\%               & 0.63$\times$          \\
    \hline
    NMT\textsubscript{L5}                                & -                  & 0.63$\times$             & 0.63$\times$             & 0\%               & 0.63$\times$          \\
    \hline
    BigLSTM                                 & -                  & 0.58$\times$             & 0.61$\times$             & 0\%               & 0.76$\times$          \\
    \hline
    LSTM-2048                               & -                  & 0.58$\times$             & 0.62$\times$             & 0\%               & 0.84$\times$          \\
    \hline
    Vgg16                                   & 0.84$\times$       & 0.75$\times$             & 0.37$\times$             & 1.96\%            & 0.69$\times$          \\
    \hline
    Vgg19                                   & 0.85$\times$       & 0.75$\times$             & 0.43$\times$             & 1.71\%            & 0.71$\times$          \\
    \hline
\end{tabular}

    }
    \vspace{-0.4cm}
\end{table}

\subsection{Evaluation of Optimizations}\label{sec:result-optimizations}

Table~\ref{tab:optimizations} shows an evaluation of various optimizations described throughout the paper.
\textbf{\textit{Input shuffling}} reduces energy consumed by data movement within a core by leveraging input reuse in CNNs.
\textbf{\textit{Shared memory sizing}} keeps the shared memory small by leveraging inter-core/tile pipelining.
The baseline here, sizes the shared memory with what would be needed without pipelining, which is 1$\times$, 50.51$\times$, 21.61$\times$, and 15.91$\times$ larger for MLPs, Deep LSTMs, Wide LSTMs, and CNNs respectively.
Note that MLP does not benefit from inter-tile pipelining because it does not exhibit any weight reuse (Section~\ref{sec:mlp}).
\textbf{\textit{Graph partitioning}} (compared to a baseline that partitions the graph randomly) reduces the number of loads, stores, sends, and receives, hence the overall energy.
\textbf{\textit{Register pressure}} is kept low by the compiler with little or no accesses from spilled registers across benchmarks.
\textbf{\textit{MVM coalescing}} runs MVMUs in parallel within a core which reduces latency.

\subsection{Design Space Exploration}\label{sec:results-sweep}

\Fig{fig:design-space} shows a PUMA tile's peak area and power efficiency swept across multiple design space parameters.
For each sweep, all other parameters are kept at the sweetspot (PUMA configuration with maximum efficiency).
Efficiency is measured using a synthetic benchmark: an MVM operation on each MVMU, followed by a VFU operation, then a ROM-Embedded RAM look-up.

Increasing the \textit{\textbf{MVMU dimension}} increases the number of crossbar multiply-add operations quadratically and the number of peripherals linearly resulting in more amortization of overhead from peripherals.
However, larger MVMUs also require ADCs with higher resolution and ADC overhead grows non-linearly with resolution, which counter-balances the amortization.
Increasing the \textit{\textbf{\# MVMUs per core}} increases efficiency because of the high efficiency of memristor crossbars relative to CMOS digital components.
However, with too many MVMUs, the VFU becomes a bottleneck which degrades efficiency.
Increasing the \textit{\textbf{VFU width}} degrades efficiency because of the low efficiency of CMOS relative to memristor crossbars.
However, a VFU that is too narrow becomes a bottleneck.
The sweetspot is found at 4 vector lanes.
Increasing the \textit{\textbf{\# cores per tile}} improves efficiency until shared memory bandwidth becomes the bottleneck.
Increasing the \textit{\textbf{register file size}} results in lower efficiency, however a register file that is too small results in too many register spills.

\begin{figure}[t]
  \centering
  \includegraphics[width=0.8\columnwidth]{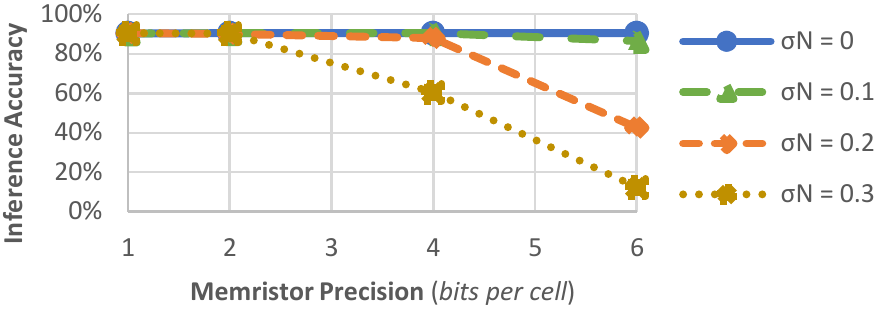}
  \vspace{-0.3cm}
  \caption{Inference Accuracy}\label{fig:results-inference-accuracy}
  \vspace{-0.3cm}
\end{figure}

\Fig{fig:results-inference-accuracy} shows PUMA's inference accuracy for different memristor bit precision (bits per device) and write noise levels ($\sigma^N$).
Higher precision can lead to larger accuracy loss due to the reduction in noise margin.
It can be seen that PUMA with 2-bit memristor performs well even at high noise levels.
Real CMOS hardware follows the $\sigma^N=0$ noise level.
Further, recent research have explored coding schemes for reliable memristor computation at high precision~\cite{feinberg2018making,roth2017fault}.

\section{Related Work}\label{sec:related}

Sze et al.~\cite{sze2017efficient} provide a thorough survey of deep learning accelerators.
In the digital realm, accelerators can be classified as
    weight stationary spatial architectures~\cite{Cavigelli:2015:OCN:2742060.2743766,Chakradhar:2010:DCC:1815961.1815993,farabetneuflow2010,gokhale2014240,park20154,sankaradas2009massively},
    output stationary spatial architectures~\cite{Du:2015:SSV:2749469.2750389,gupta2015deep,peemen2013memory},
    spatial architectures with no local reuse~\cite{chenDianNao2014,chen2014dadiannao,zhangfpga2015},
    and row stationary spatial architectures~\cite{chen2016eyeriss}.
Many designs also support optimizations and features including
    weight pruning and exploiting sparsity~\cite{Albericio:2017:BDN:3123939.3123982,albericio2016cnvlutin,chung2016simplifying,Ding:2017:CIC:3123939.3124552,han2016eie,Parashar:2017:SAC:3079856.3080254,reagen2016minerva,zhang2016cambricon},
    reducing precision~\cite{andri2017yodann,judd2016stripes},
    stochastic computing~\cite{kim2016dynamic,ren2017sc},
    layer fusing~\cite{alwani2016fused},
    meeting QoS/QoR requirements~\cite{wang2017real},
    graph tuning~\cite{Ji2018},
    and reconfigurable interconnects~\cite{Kwon2018}.
Digital accelerators have varied in their degree of flexibility, ranging from
    custom accelerators specialized for a particular field~\cite{Cai2018,murray2016microarchitecture,Song2018,yazdani2016ultra},
    to accelerators that are fully programmable via an ISA~\cite{Jouppi:2017:IPA:3079856.3080246,Liu:2015:PPM:2694344.2694358,liu2016cambricon,Venkataramani:2017:SSC:3079856.3080244,zhang2016cambricon}.
FPGAs have also been popular targets for building accelerators \cite{farabet2009cnp,guo2017survey,mahajan2016tabla,peemen2013memory,sharma2016high,Shen:2017:MCA:3079856.3080221,wang2016deepburning,zhangfpga2015}.
All these works remain in the digital domain, while PUMA leverages hybrid digital-analog computing.

Near-memory acceleration for ML has been proposed using
    DRAM~\cite{gao2017tetris,kim2016neurocube,li2017drisa}
    and SRAM~\cite{wang2014error,zhang2016machine}.
PUMA uses non-volatile memristive crossbars for near-memory acceleration.

Many machine learning accelerators have been proposed that leverage memristor crossbars~\cite{ankit2017resparc,ankit2017trannsformer,bojnordi2016memristive,cheng2017time,chi2016prime,Hu:2012:HRB:2228360.2228448,ji2016neutrams,Kim:2015:RDN:2767119.2700234,liu2015reno,ramasubramanian2014spindle,shafiee2016isaac,song2017pipelayer}.
These accelerators have been demonstrated on several types of workloads including
    BSBs~\cite{Hu:2012:HRB:2228360.2228448},
    MLPs~\cite{cheng2017time,Feinberg2018,liu2015reno,ramasubramanian2014spindle},
    SNNs~\cite{ankit2017resparc,Kim:2015:RDN:2767119.2700234},
    BMs~\cite{bojnordi2016memristive},
    and CNNs~\cite{cheng2017time,chi2016prime,shafiee2016isaac,song2017pipelayer,wang2017group}.
Some accelerators
    support inference only~\cite{ankit2017resparc,chi2016prime,Hu:2012:HRB:2228360.2228448,liu2015reno,ramasubramanian2014spindle,shafiee2016isaac}
    while others also support training~\cite{bojnordi2016memristive,cheng2017time,Kim:2015:RDN:2767119.2700234,song2017pipelayer}.
Ji et al.~\cite{ji2016neutrams,ji2018bridge} transforms neural networks to configure such accelerators.
These accelerators vary in flexibility, but even the most flexible rely on state machine configuration and have only been demonstrated on a few types of workloads.
PUMA is the first memristor-based accelerator for machine learning inference that is ISA-programmable and general-purpose.

Fujiki et al.~\cite{fujiki2018memory} propose an ISA-programmable memristor-based accelerator.
Their accelerator is a data-parallel accelerator whereas PUMA is a data-flow accelerator with more capability for producer-consumer synchronization.
Moreover, their accelerator optimizes crossbars for vector operations in general-purpose workloads whereas PUMA optimizes crossbars for MVM operations prevalent in machine learning and uses digital VFUs for vector operations rather than crossbars.

Chung et al.~\cite{chung2018serving} propose Brainwave, which is a spatial accelerator built with FPGAs.
Compared to Brainwave, a PUMA core performs 0.26 million 16-bit ops, equivalent to 1.04 million 8-bit ops, per coalesced MVM instruction.
A Brainwave NPU performs 1.3million 8-bit ops per instruction.
Therefore, PUMA and Brainwave have comparable control granularity while PUMA has 40.8x higher storage-density (Brainwave Stratix10 estimate).

Memristors have also been proposed for building byte-addressable non-volatile memories.
There have been various works centered around system support for non-volatile memory, including
    file systems~\cite{condit2009better},
    memory allocators~\cite{Bhandari:2016:MFR:2983990.2984019,oukid2017memory},
    programming models~\cite{Chakrabarti:2014:ALL:2660193.2660224,Coburn:2011:NMP:1950365.1950380,Volos:2011:MLP:1950365.1950379},
    durable data structures~\cite{debnath2016revisiting,izraelevitz2016linearizability,yang2015nv},
    representation of pointers~\cite{chen2017efficient,Cohen:2018:ORN:3288538.3276523,el2017savi,ElHajj:2016:SPM:2872362.2872366},
    and
    architecture support~\cite{joshi2017atom,nalli2017analysis,ogleari2018steal,shin2017proteus}.

There have been concerns in the community about memristor manufacturability.
We distinguish between medium-density embedded memristor applications and high-density storage-class memory (SCM).
Memristors in PUMA use 1T1R configuration which have been shown to have good manufacturability~\cite{hayakawa2015highly}.
They are very different from SCM, where the selector transistor may be replaced with an in-line two-terminal selector device for higher density which complicates manufacturability.
Panasonic formed a joined venture with UMC foundry in 2017 to enable integration of memristors to UMC 40nm CMOS process with first samples planned in 2018~\cite{panasonic2017}.
TSMC also completed development of their own memristor technology that entered risk production in 40nm ULP CMOS process node at the end of 2017~\cite{tsmc2017}.

\section{Conclusion}\label{sec:conclusion}

PUMA is the first ISA-programmable accelerator for ML inference that uses hybrid CMOS-memristor technology.
It enhances memristor crossbars with general purpose execution units carefully designed to maintain crossbar area/energy efficiency and storage density.
Our accelerator design comes with
    a complete compiler to transform high-level code to PUMA ISA
    and a detailed simulator for estimating performance and energy consumption.
Our evaluations show that PUMA can achieve significant improvements compared to state-of-the-art CPUs, GPUs, and ASICs for ML acceleration.

\begin{acks}

This work is supported by Hewlett Packard Labs and the US Department of Energy (DOE) under Cooperative Agreement DE-SC0012199, the Blackcomb 2 Project. 
In addition, John Paul Strachan acknowledges support in part from the Intelligence Advanced Research Projects Activity (IARPA) via contract number 2017-17013000002.
This work was also supported in part by the Center for Brain-inspired Computing (C-BRIC), one of six centers in JUMP, a DARPA sponsored Semiconductor Research Corporation (SRC) program.

\end{acks}

\clearpage

%
% The next two lines define the bibliography style to be used, and the bibliography file.
\bibliographystyle{ACM-Reference-Format}
\bibliography{ref}

\end{document}